\documentclass{sig-alternate-sigmod08}

\usepackage{PageSetting}
\usepackage{wrapfig,epsfig}

\usepackage{algorithm}
\usepackage{algorithmic}

\def\done{\hspace*{\fill} $\framebox[2mm]{}$ \medskip}

 \newcommand{\QED}{%
   \ifmmode 
   \else \leavevmode\unskip\penalty9999 \hbox{}\nobreak\hfill
   \fi
   \quad\mbox{\rule[0pt]{1.5ex}{1.5ex}}}

\title{Preserving Individual Privacy in Serial Data Publishing}

\author{
  Raymond Chi-Wing Wong$^1$, Ada Wai-Chee Fu$^2$, Jia Liu$^2$,
  Ke Wang$^3$, Yabo Xu$^3$\vspace*{0.4cm} \\
\begin{tabular}{c c c}
$^1$ Hong Kong University of Science and Technology & $^2$Chinese University of Hong Kong \\
\normalsize{raywong@cse.ust.hk} & \normalsize{\{adafu,jliu\}@cse.cuhk.edu.hk} \\
 \multicolumn{2}{c}{$^3$ Simon Fraser University }\\
\multicolumn{2}{c}{\normalsize{\{wangk,yxu\}@cs.sfu.ca}}
\end{tabular}
}
\date{ }

\begin{document}

\maketitle

\begin{sloppy}

\begin{abstract}
While previous works on privacy-preserving serial data publishing
consider the scenario where sensitive values may persist over
multiple data releases, we find that no previous work has sufficient
protection provided for sensitive values that can change over time,
which should be the more common case. In this work, we propose to
study the privacy guarantee for such transient sensitive values,
which we call the \emph{global guarantee}. We formally define the
problem for achieving this guarantee and derive some theoretical
properties for this problem. We show that the anonymized group sizes
used in the data anonymization is a key factor in protecting
individual privacy in serial publication. We propose two strategies
for anonymization targeting at minimizing the average group size and
the maximum group size. Finally, we conduct experiments on a medical
dataset to show that our method is highly efficient and also
produces published data of very high utility.
\end{abstract}

\section{Introduction}
\label{sec:intro}

Recently, there has been much study on the issues in
privacy-preserving data publishing
\cite{AggarwalICDT05,Incognito,multidimensional-Kanonymity,BOY+05,l-diversity,WLFW-kdd06,DXTZZ07,LL07,ZKS+07,GTK08,TXLZ08,LL08}.
Most previous works deal with privacy protection when only one
instance of the data is published. However, in many applications,
data is published at regular time intervals. For example, the
medical data from a hospital may be published twice a year. Some
recent papers \cite{PXW+07,XT07,FWFP+08,ByunSBL06,WF+06,BFW+08}
study the privacy protection issues for \emph{multiple} data
publications of multiple instances of the data. We refer to such
data publishing \emph{serial data publishing}.

Following the settings of previous works, we assume that there is a
sensitive attribute which contains sensitive values that should not
be linked to the individuals in the database. A common example of
such a sensitive attribute is diseases. While some diseases such as
flu or stomach virus may not be very sensitive, some diseases such
as chlamydia (a sex disease) can be considered highly sensitive. In
serial publishing of such a set of data, the disease values attached
to a certain individual can change over time.

A typical guarantee we want to achieve is that the probability that
an adversary can derive for the linkage of a person to a sensitive
value is no more than $1/\ell$. This is well-known to be a simple
form of $\ell$-diversity \cite{l-diversity}. This guarantee sounds
innocent enough for a single release data publication. However, when
it comes to serial data publishing, the objective becomes quite
illusive and requires a much closer look. In serial publishing, the
individuals that are recorded in the data may change, and the
sensitive values related to individuals may also change. We assume
that the sensitive values can change freely.

Let us consider a sensitive disease chlamydia, which is a sex
disease that is easily curable. Suppose that there exist 3 records
of an individual $o$ in 3 different medical data releases. It is
obvious that typically $o$ would not want anyone to deduce with high
confidence from these released data that s/he has ever contracted
chlamydia in the past. Here, the past practically corresponds to
\emph{one or more} of the three data releases. Therefore, if from
these data releases, an adversary can deduce with high confidence
that $o$ has contracted chlamydia in one or more of the three
releases, privacy would have been breached. To protect privacy, we
would like the probability of any individual being linked to a
sensitive value in one or more data releases to be bounded from the
above by $1/\ell$. Let us call this privacy guarantee the
\emph{global guarantee} and the value $1/\ell$ the \emph{privacy
threshold}.

Though the global guarantee requirement seems to be quite obvious,
to the best of our knowledge, no existing work has considered such a
guarantee. Instead, the closest guarantee of previous works is the
following: for \emph{each} of the data releases, $o$ can be linked
to chlamydia with a probability of no more than $1/\ell$.
Let us call this guarantee the \emph{localized guarantee}. Would
this guarantee be equivalent to the above global guarantee ? In
order to answer this question, let us look at an example.

Consider two raw medical tables (or micro data) $T_1$ and $T_2$ as
shown in Figure~\ref{tab:introExample1} at time points 1 and 2,
respectively. Suppose that they contain records for the individuals
$o_1, o_2, o_3, o_4,o_5$. There are two kinds of attributes, namely
\emph{quasi-identifier (QID)} attributes and \emph{sensitive}
attributes. Quasi-identifier attributes are attributes that can be
used to identify an individual with the help of an external source
such as a voter registration list
\cite{sweeney-kanonymity-model,multidimensional-Kanonymity,Incognito,XT06a}.
In this example, sex and zipcode are the quasi-identifier
attributes, while disease is the sensitive attribute. Attribute id
is used for illustration purpose and does not appear in the
published table. We assume that each individual owns at most one
tuple in each table at each time point. Furthermore, we assume no
additional background knowledge about the linkage of individuals to
diseases, and the sensitive values linked to individuals can be
freely updated from one release to the next release.

\begin{figure}[tbph]
\scriptsize \hspace*{-3mm}
\begin{tabular}{c c}
\begin{minipage}[htbp]{4cm}
\center
\begin{tabular}{| c | c | c| c |}\hline
 Id & Sex & Zip- & Disease \\
 && code & \\ \hline
 $o_1$ &  M & 65001 & flu\\\hline
 $o_2$ & M & 65002 & chlamydia\\\hline
 $o_3$ & F & 65014 & flu \\\hline
 $o_4$ & F & 65015 & fever \\\hline
\end{tabular}
\end{minipage}
&
\begin{minipage}[htbp]{4cm}
\center
\begin{tabular}{| c | c | c| c |}\hline
Id &  Sex & Zip- & Disease \\
 && code & \\ \hline
  $o_1$ & M & 65001 & chlamydia\\\hline
$o_2$ &  M & 65002 & flu \\\hline
 $o_3$ &  F & 65014 & fever\\\hline
$o_5$ &  F & 65010 & flu \\\hline
\end{tabular}
\end{minipage}
\\
(a) $T_1$ & (b) $T_2$
\end{tabular}
\caption{A motivating example} \label{tab:introExample1}
\end{figure}

\begin{figure}[hb]
\scriptsize \center
\begin{tabular}{c c}
\begin{minipage}[htbp]{4cm}
\center
\begin{tabular}{| c | c| c |}\hline
  Sex & Zipcode & Disease \\ \hline
  M & 6500* & flu\\\hline
  M & 6500* & chlamydia\\\hline
  F & 6501* & flu \\\hline
  F & 6501* & fever \\\hline
\end{tabular}
\end{minipage}
&
\begin{minipage}[htbp]{4cm}
\center
\begin{tabular}{| c | c| c |}\hline
  Sex & Zipcode & Disease \\ \hline
  M & 6500* & chlamydia\\\hline
  M & 6500* & flu\\\hline
  F & 6501* & fever \\\hline
  F & 6501* & flu \\\hline
\end{tabular}
\end{minipage}
\\
(a) $T_1^*$ & (b) $T_2^*$
\end{tabular}
\caption{Anonymization for $T_1$ and $T_2$}
\label{tab:introExample2}
\end{figure}

\begin{figure*}[htbp]
\scriptsize
\center
\begin{tabular}{c c c c}
\begin{minipage}[htbp]{4.1cm}
\center
\begin{tabular}{| c | c| c |}\hline
  Sex & Zipcode & Disease \\ \hline
  M & 65001 & flu\\\hline
  M & 65002 & chlamydia\\\hline
\end{tabular}\\
$T_1$\medskip\\
\begin{tabular}{| c | c| c |}\hline
  Sex & Zipcode & Disease \\ \hline
  M & 65001 & flu\\\hline
  M & 65002 & chlamydia\\\hline
\end{tabular}\\
$T_2$\\
\end{minipage}
&
\begin{minipage}[htbp]{4.1cm}
\center
\begin{tabular}{| c | c| c |}\hline
  Sex & Zipcode & Disease \\ \hline
  M & 65001 & flu\\\hline
  M & 65002 & chlamydia\\\hline
\end{tabular}\\
$T_1$\medskip\\
\begin{tabular}{| c | c| c |}\hline
  Sex & Zipcode & Disease \\ \hline
  M & 65001 & chlamydia\\\hline
  M & 65002 & flu \\\hline
\end{tabular}\\
$T_2$\\
\end{minipage}
&
\begin{minipage}[htbp]{4.1cm}
\center
\begin{tabular}{| c | c| c |}\hline
  Sex & Zipcode & Disease \\ \hline
  M & 65001 & chlamydia\\\hline
  M & 65002 & flu \\\hline
\end{tabular}\\
$T_1$\medskip\\
\begin{tabular}{| c | c| c |}\hline
  Sex & Zipcode & Disease \\ \hline
  M & 65001 & flu \\\hline
  M & 65002 & chlamydia\\\hline
\end{tabular}\\
$T_2$\\
\end{minipage}
&
\begin{minipage}[htbp]{4.1cm}
\center
\begin{tabular}{| c | c| c |}\hline
  Sex & Zipcode & Disease \\ \hline
  M & 65001 & chlamydia\\\hline
  M & 65002 & flu \\\hline
\end{tabular}\\
$T_1$\medskip\\
\begin{tabular}{| c | c| c |}\hline
  Sex & Zipcode & Disease \\ \hline
  M & 65001 & chlamydia \\\hline
  M & 65002 & flu \\\hline
\end{tabular}\\
$T_2$\\
\end{minipage}
\\
(a) Possible world 1 $w_1$ & (b) Possible world 2 $w_2$  & (c) Possible world 3 $w_3$  & (d) Possible world 4 $w_4$
\end{tabular}
\caption{Possible worlds for $G_1$ and $G_2$}
\label{tab:possWorldExample}
\end{figure*}

Assume that the privacy threshold is $1/\ell = 1/2$. In a typical
data anonymization
\cite{sweeney-kanonymity-model,multidimensional-Kanonymity,Incognito,XT06a},
in order to protect individual privacy, the QID attributes of the
raw table are \emph{generalized} or \emph{bucketized} in order to
form some \emph{anonymized groups} ($\cal{AG}$) to hide the linkage
between an individual and a sensitive value. For example, table
$T_1^*$ in Figure~\ref{tab:introExample2}(a) is a \emph{generalized}
table of $T_1$ in Figure~\ref{tab:introExample1}. We generalize the
zip code of the first two tuples to 6500* so that they have the same
QID values in $T_1^*$. We say that these two tuples form an
\emph{anonymized group}. It is easy to see that in each published
table $T_1^*$ or $T_2^*$, the probability of linking any individual
to chlamydia or flu is at most 1/2, which satisfies the localized
guarantee. The question is whether this satisfies the global privacy
guarantee with a threshold of $1/2$.

For the sake of illustration, let us focus on the anonymized groups
$G_1$ and $G_2$ containing the first two tuples in tables $T_1^*$
and $T_2^*$ in Figure~\ref{tab:introExample2}, respectively. The
probability in serial publishing can be derived by the possible
world analysis. There are four possible \emph{worlds} for $G_1$ and
$G_2$ in these two published tables, as shown in
Figure~\ref{tab:possWorldExample}. Here each \emph{possible world}
is one possible way to assign the diseases to the individuals in
such a way that is consistent with the published tables. Therefore,
each possible world is a possible assignment of the sensitive values
to the individuals at all the publication time points for groups
$G_1$ and $G_2$. Note that an individual can be assigned to
different values at different data releases, and the assignment in
one data release is independent of the assignment in another
release.

Consider individual $o_2$. Among the four possible worlds, three
possible worlds link $o_2$ to ``chlamydia", namely $w_1, w_2$ and
$w_3$. In $w_1$ and $w_2$, the linkage occurs at $T_1$, and in
$w_3$, the linkage occurs at $T_2$. Thus, the probability that $o_2$
is linked to ``chlamydia" in at least one of the tables is equal to
$3/4$, which is greater than $1/2$, the intended privacy threshold.
From this example, we can see that localized guarantee does not
imply global guarantee.

\begin{figure}[htbp]
\scriptsize \center
\begin{tabular}{c c}
\begin{minipage}[htbp]{4cm}
\center
\begin{tabular}{| c | c| c |}\hline
  Sex & Zipcode & Disease \\ \hline
  M/F & 650** & flu \\\hline
  M/F & 650** & chlamydia\\\hline
  M/F & 650** & flu \\\hline
  M/F & 650** & fever \\\hline
\end{tabular}
\end{minipage}
&
\begin{minipage}[htbp]{4cm}
\center
\begin{tabular}{| c | c| c |}\hline
  Sex & Zipcode & Disease \\ \hline
  M/F & 650** & chlamydia\\\hline
  M/F & 650** & flu\\\hline
  M/F & 650** & fever \\\hline
  M/F & 650** & flu \\\hline
\end{tabular}
\end{minipage}
\\
(a) $T_1*$ & (b) $T_2*$
\end{tabular}
\caption{Anonymization for global guarantee} \label{size4}
\end{figure}


In this paper, we show that in order to ensure the global guarantee,
the sizes of the anonymized groups need to be bigger than that
needed for localized guarantee. In the above example, we can use
size 4 anonymized groups as shown in Figure \ref{size4}. There will
be $4! \times 4!$ possible worlds. It is easy to see that $3/4$ of
the possible worlds do not assign chlamydia to $o_2$ in the first
release, $3/4$ of them do not assign chlamydia to $o_2$ in the
second release, and $3/4 \times 3/4 = 9/16$ of the possible worlds
do not assign chlamydia to $o_2$ in both releases. The remaining
possible worlds assign chlamydia to $o_2$ in at least one of the two
releases. Hence, the privacy breach probability = $1 - 9/16 = 7/16 <
1/2$.

The contributions of this paper include the following: We point out
the problem of privacy breach that arises with localized guarantee
and propose to study the problem of global guarantee in privacy
preserving serial data publishing. We formally analyze the privacy
breach with transient sensitive values. Useful properties related to
the anonymization under the global guarantee are derived. These
properties are related to the anonymized group sizes. Typically
group sizes greater than that required for the localized guarantee
will be needed to attain the global guarantee. These properties are
then leveraged in the proposal of new anonymization strategies that
can minimize the information loss. We have also conducted extensive
experiments with a real medical dataset to verify our techniques.
The results show that our methodology are very promising in real
world applications.

The rest of this paper is organized as follows.
Section~\ref{sec:related} surveys the previous related works.
Section~\ref{sec:probDef} contains our problem definition.
Section~\ref{sec:formula} describes a general formula for the breach
probability. Section~\ref{sec:properties} discusses some key
properties for this problem. Section~\ref{sec:alg} describes our
methodology for privacy protection. Section~\ref{sec:implement}
suggests a possible implementation. Section~\ref{sec:exp} is an
empirical study. 
Section~\ref{sec:concl} concludes our work and points out some
possible future directions.

\section{Related Work}
\label{sec:related}

Here, we summarize the previous works on the problem of privacy
preserving serial data publishing. $k$-anonymity has been considered
in \cite{FWFP+08} and \cite{PXW+07} for serial publication allowing
only insertions, but they do not consider the linkage probabilities
to sensitive values. The work in \cite{WF+06} considers sequential
releases for different attribute subsets for the same dataset, which
is different from our definition of serial publishing.

There are some more related works that attempt to avoid the linkage
of individuals to sensitive values. Delay publising is proposed in
\cite{ByunSBL06} to avoid problems of insertions, but deletion and
updates are not considered. While \cite{XT07} considers both
insertions and deletions, both \cite{ByunSBL06} and \cite{XT07} make
the assumption that when an individual appears in consecutive data
releases, then the sensitive value for that individual is not
changed. As pointed out in \cite{BFW+08}, this assumption is not
realistic. Also the protection in \cite{XT07} is record-based and
not individual-based. This is quite problematic, as in our running
examples, there are two \emph{records} for one \emph{individual}
$o_2$, namely, $t_1$ in table $T_1$ and $t_2$ in table $T_2$ (note
that $T_1$ and $T_2$ need not be consecutive releases, so that the
sensitive value linked to $o_2$ can change even if we adopt the
above unrealistic assumption in \cite{ByunSBL06,XT07}). If we
consider just tuple $t_1$, then there are only 2 possible worlds
where $t_1$ is linked to chlamydia in Figure
\ref{tab:possWorldExample}, namely $w_1$ and $w_2$. If we just
consider tuple $t_2$, there are also only 2 possible worlds linking
it to chlamydia, namely $w_1$ and $w_3$. Hence, $T_1^*$ and $T_2^*$
satisfy the record-based requirement of \cite{XT07} if the risk
threshold is 0.5. In fact, these are possible tables generated by
the mechanism proposed in \cite{XT07}. However, we have shown that
this anonymization does not provide the expected protection for the
individuals.

The $\ell$-scarcity model is introduced in \cite{BFW+08} to handle
the situations when some data may be permanent so that once an
individual is linked to such a value, the linkage will remain in
subsequent releases whenever the individual appears (not limited to
consecutive releases only). However, for transient sensitive values,
\cite{XT07} and \cite{BFW+08} adopt the following principle.

\begin{principle}[Localized Guarantee]
For each release of the data publication, the probability that an
individual is linked to a sensitive value is bounded by a threshold.
\label{principle:everyPrivacyProtection}
\end{principle}

However, we have seen in the example in the previous section that
this cannot satisfy the expected privacy requirement. Hence, we
consider the following principle.

\begin{principle}[Global Guarantee]
Over all the published releases, the probability that an individual
has ever been linked to a sensitive value is bounded by a threshold.
\label{principle:atLeastOnePrivacyProtection}
\end{principle}

Although the privacy guarantee is the most important data
publication criterion, the published data must also provide a
reasonable level of utility so that it can be useful for
applications such as data mining or data analysis. Utility is a
tradeoff for the privacy guarantee since anonymization of data
introduces information loss. There are different definitions of
utility in the existing literature. Here, we briefly describe some
common definitions.

The anonymized group sizes have been considered in utility metrics.
The average group size is considered in \cite{l-diversity}. In
\cite{Bayardo-optimal}, the discernability model assigns a penalty
to each tuple $t$ as determined by the square of the size of the
anonymized group for $t$. In \cite{multidimensional-Kanonymity}, the
normalized average anonymized group size metric is proposed, which
is given by the total number of tuples in the table divided by the
product of the total number of anonymized groups and a value $k$ (for $k$-anonymity). Here,
the best case occurs when each group has size $k$.

Other works \cite{XT06a,XWP-kdd06,WFW+07} consider categorical data
that comes with a taxonomy so that the information loss is measured
with respective to the structure in the taxonomy  when data are
generalized from the leaf nodes to higher levels in the taxonomy.
Both \cite{KG06} and \cite{XT06b} measure utility by comparing the
data distributions before and after anonymization. Recently,
\cite{RSH07} and \cite{ZKS+07} consider the accuracy in answering
aggregate queries to be a measure of utility.

\cite{Iyengar-kdd02,WangKe-template,EDBT04} assume that the data is
utilized for classification and hence define the utility accordingly. The anonymization mechanisms in
\cite{Meyerson-pods04,AggarwalICDT05,XWFY08} are by means of
suppressing data entries in the table, and hence information loss is
measured by the number of suppressed entries.

\section{Problem Definition}
\label{sec:probDef}

Suppose tables $T_1, T_2, ..., T_k$ are generated at time points,
$1, 2, ..., k$, respectively. Each table $T_i$ has two kinds of
attributes, \emph{quasi-identifier attributes} and \emph{sensitive
attributes}. For the sake of illustration, we consider one single
sensitive attribute $S$ containing $|S|$ values, namely $s_1, s_2,
..., s_{|S|}$. Assume that the sensitive values for individuals can
freely change from one release to another release so that the
linkage of an individual $o$ to a sensitive value $s$ in one data
release has no effect on the linkage of $o$ to any other sensitive
value in any other data release. Assume at each time point $j$, a
data publisher generates an anonymized version $T_j^*$ of $T_j$ for
data publishing so that each record in $T_j$ will belong to one
anonymized group $G$ in $T_j^*$. Given an anonymized group $G$, we
define $G.S$ to be a multi-set containing all sensitive values in
$G$, and $G.I$ to be the set of individuals that appear in $G$.

\begin{definition}[Possible World] A series of tables
 $TS = \{T_1^p, T_2^p,..., T_k^p\}$ is a possible
world for published tables $\{T_1^*, T_2^*,..., T_k^*\}$
 if the following requirement is satisfied. For each $i \in [1, k]$,
\begin{enumerate}
  \item
    there is a one-to-one corresponding between individuals in $T_i^p$ and
    individuals in $T_i^*$
  \item for each anonymized group $G$ in $T_i^*$,
          the multi-set of the sensitive values of the corresponding individuals in $T_i^p$
          is equal to $G.S$.
\end{enumerate}
\end{definition}


Let $p(o, s, k)$ be the probability that an individual $o$ is linked
to $s$ in at least one published table among published tables
$T_1^*, T_2^*,..., T_k^*$.

Let $t.S$ stand for the sensitive value of tuple $t$. We say that
$o$ is linked to $s$ in a table $T_i^p$ if for the tuple $t$ of $o$
in $T_i^p$, $t.S=s$. Following previous works, we define the
probability based on the \emph{possible worlds} as follows.

\begin{definition}[Breach Probability]
The breach probability is given by
\begin{eqnarray}
p(o,s,k) = \frac{W_{link}(o, s, k)}{W_{total,k}}
\label{eqn:breachProb}
\end{eqnarray}
where $W_{link}(o, s, k)$ is the total number of possible worlds
where $o$ is linked to $s$ in at least one published table among
$T_1^p, T_2^p, ..., T_k^p$ and $W_{total,k}$ is the total number of
possible worlds for published tables $T_1^*, T_2^*, ..., T_k^*$.
\label{def:breachProb}
\end{definition}

 We will describe how
we derive a general formula to calculate $p(o, s, k)$ in
Section~\ref{sec:formula}.

While privacy breach is the most important concern, the utility of
the published data also need to be preserved. There are different
definitions of utility in the existing literature. Some commonly
adopted utility measurements are described in Section
\ref{sec:related}.

In this paper, we are studying the following problem.

\begin{problem}
Given a privacy parameter $\ell$ (a positive integer), a utility
measurement, $k-1$ published tables, namely $T_1^*, T_2^*,...,
T_{k-1}^*$ and one raw table $T_k$, we want to generate a published
table $T_{k}^*$ from $T_k$ such that the utility is maximized, and
for each individual $o$ and each sensitive value $s$,
$$
  p(o, s, k) \le 1/\ell
$$
\end{problem}

Note that the above problem definition follows
Principle~\ref{principle:atLeastOnePrivacyProtection} for global
guarantee as discussed in Section~\ref{sec:related}.

\subsection{Global versus Localized Guarantee}
\label{subsec:strongerPrivacyRequirement}

Here, we show that protecting individual privacy with
Principle~\ref{principle:atLeastOnePrivacyProtection} (global
guarantee) implies protecting individual privacy with
Principle~\ref{principle:everyPrivacyProtection} (localized
guarantee). Under Principle~\ref{principle:everyPrivacyProtection},
let $q(o,s,j,k)$ be the probability that an individual $o$ is linked
to a sensitive value $s$ in the $j$-th table. Following the
definition of probability adopted in most previous works
\cite{XT07,BFW+08}, we have
$$
q(o,s,j,k) = \frac{{L}_{link}(o, s, j, k)}{W_{total,k}}
$$
where ${L}_{link}(o, s, j, k)$ is the total number of possible
worlds in which
 $o$ is linked to $s$ in the $j$-th table and $W_{total,k}$ is the total number of
possible worlds for the $k$ published tables.

In our running example, $k$=2 and from
Figure~\ref{tab:possWorldExample}, there are four possible worlds,
$W_{total,k} =$ 4. Consider published table $T_1^*$. There are two
possible worlds where $o_2$ is linked to chlamydia ($s$), namely
$w_1$ and $w_2$. Thus, ${L}_{link}(o_2, s, 1, k) = 2$ and $
q(o_2,s,1,k) = \frac{2}{4} = \frac{1}{2} $. Similarly, when $j = 2$,
$q(o_2,s,2,k) = \frac{1}{2}$.

%

In general, it is obvious that $W_{link}(o, s, k) \ge L_{link}(o, s,
j, k)$ for any $j \in [1, k]$. We derive that
$$
p(o,s,k) \ge q(o,s,j,k)
$$

Hence we have the following lemma.

\begin{lemma}
If $p(o, s, k) \le 1/\ell$ (under
Principle~\ref{principle:atLeastOnePrivacyProtection}), then for any
$j \in [1, k]$, $q(o, s, j, k) \le 1/\ell$ (under
Principle~\ref{principle:everyPrivacyProtection}).
\end{lemma}



\begin{corollary}
Principle~\ref{principle:atLeastOnePrivacyProtection} (global
guarantee)  is a strictly stronger requirement than
Principle~\ref{principle:everyPrivacyProtection} (localized
guarantee).
\end{corollary}

%
%

\section{Breach Probability Analysis}
\label{sec:formula}


In this section, we consider how the breach probability $p(o,s,k)$
can be derived. For privacy breach, we focus on the possible
assignment of sensitive values to one individual at a time.
Therefore, we introduce the following possible world definition to
deal with assignments to a particular individual.

\begin{definition}[$\mathcal{AG}_i$]
At any data release, let ${\cal AG}_i(o)$ be the anonymized group that
contains the record for individual $o$ in published table $T_i^*$. \end{definition}

For the sake of clarity, if the context is clear, we omit
the subscript and denote ${\cal AG}_i(o)$ by  ${\cal AG}(o)$.

\begin{definition}[Possible World for $o$]
Given a possible world $TS = \{T_1^p, T_2^p,..., T_k^p\}$ for
$\{T_1^*, T_2^*,..., T_k^*\}$. Let us extract the tuples in each
$T_i^p$ that correspond to the tuples in the anonymized group ${\cal
AG}_i(o)$ (containing individual $o$ in $T_i^*$) to form table
$T_i^p(o)$. Then, the series of smaller tables, denoted by $TS(o)$
which is equal to $\{T_1^p(o), T_2^p(o),..., T_k^p(o)\}$,
form a possible world for ${\cal AG}_1(o)$, ... ${\cal AG}_k(o)$. We
also say that that $TS(o)$ is a possible world for $o$ for $\{T_1^*,
T_2^*,..., T_k^*\}$.
\end{definition}

 For example,
Figure~\ref{tab:possWorldExample} shows all the possible worlds for
$G_1$ and $G_2$ for $o_2$ in the published tables shown in
Figure~\ref{tab:introExample2}(a) and
Figure~\ref{tab:introExample2}(b). Note that in the above
definition, if $o$ does not appear in a table $T_i$, then
$T_i^p(o)$ is an empty table.

\subsection{Possible World Analysis}
\label{subsec:possWorldAnalysis}

Since the sensitive values are transient and we do not assume any
additional knowledge about the data linkage, the assignment of
sensitive values to individuals in groups other than ${\cal AG}(o)$
are independent of the assignment to the individuals in ${\cal
AG}(o)$. Hence, we arrive at the following lemma.

\begin{lemma}
The value of $p(o,s,k)$ can be derived based on the analysis of the
possible worlds for $o$.
\end{lemma}

The above lemma helps to greatly simplify the analysis of the
privacy breach by considering only ${\cal AG}(o)$ in each data
release. In the following, we may refer to a possible world for $o$
simply as a possible world.


Consider an anonymized group ${\cal AG}(o)$ in $T_j$ for individual
$o$. Let $n_j$ be the size ${\cal AG}(o)$. Let $n_{j,i}$ be the
total number of tuples in ${\cal AG}(o)$ with sensitive value $s_i$
for $i = 1, 2, ..., |S|$. The total number of possible worlds for
${\cal AG}(o)$ can be derived by \emph{combinatorial} analysis.

\begin{lemma}[No. of Poss. Worlds for Single Table]
The total number of possible worlds for the anonymized group ${\cal
AG}(o)$ in a \emph{single} published table $T_j^*$ is equal to
$$
W_j = \frac{n_j!}{\prod_{i=1}^{|S|}n_{j,i}!}
$$
\label{lemma:simplifiedSingleTable}
\end{lemma}


For example, consider an anonymized group of size 4 containing two
$s_1$ values, one $s_2$ value and one $s_3$ value in $T_j^*$. Then,
$W_j$ is equal to $\frac{4!}{2! \times 1! \times 1!} = 12$.

\subsection{Breach Probability}
\label{subsec:breachProb}

Recall that our objective is to compute $p(o, s, k)$ which involves
two major components, namely $W_{link}(o, s, k)$ and $W_{total,k}$.
In the following, we will describe how we obtain the values of these
two components.

By Lemma~\ref{lemma:simplifiedSingleTable}, 
the total number of possible worlds for $o$ in the published tables
$T_1^*, T_2^*, ..., T_k^*$, denoted by $W_{total,k}$, is equal to
\begin{eqnarray}
W_{total,k} = \prod_{j=1}^k W_j = \prod_{j=1}^{k}
\frac{n_j!}{\prod_{i=1}^{|S|}n_{j,i}!}
\label{eqnarray:totalNoOfWorldMultipleTable}
\end{eqnarray}

Next, we will describe how to obtain the formula for $W_{link}(o, s,
k)$. Without loss of generality, we consider the privacy protection
for an arbitrary sensitive value $s = s_1$.
The following analysis applies for each sensitive value.

Note that, for any arbitrary sensitive value $s_1$, we have the following.
$$
  W_{total, k} = W_{link}(o, s_1, k) + \overline{W}_{link}(o, s_1, k)
$$
where $\overline{W}_{link}(o, s_1, k)$ is the total number of possible
worlds where $o$ is not linked to $s_1$ in all $k$ published tables, namely
$T_1^p, T_2^p, ..., T_k^p$.
Thus,
$$
  W_{link}(o, s_1, k) = W_{total, k} -  \overline{W}_{link}(o, s_1, k)
$$
Next, we will show how we derive $\overline{W}_{link}(o, s_1, k)$.
Let $\theta(o, s_1, j)$ be the total number of possible worlds for table
$T_j^p$ (treated as a singleton table series) that
$o$ is not linked to $s_1$.

Consider a possible table $T_j^p$ where $o$ is not linked to $s_1$.
Since $o$ is not linked to $s_1$ in $T_j^p$,
$o$ is linked to a sensitive value $s_q$ where $q \neq 1$ in $T_j^p$.
The number of possible worlds for $T_j^p$ where $o$ is linked to
$s_q$ in $T_j^p$ is equal to
$$
W_{sq,j} = \frac{(n_j-1)!}{(n_{j,q}-1)!\prod_{i=1\mbox{ and }i\neq
q}^{|S|}n_{j,i}!}
$$
         By considering all sensitive values $s_q$ where $q \in [2, |S|]$,
         the total number of possible worlds for $T_j^p$
         where $o$ is not linked to $s_1$ (i.e., $\theta(o, s _1, j)$)
         is equal to
         \begin{eqnarray*}
              \sum_{q=2}^{|S|} W_{sq, j}   & = & \sum_{q=2}^{|S|} \frac{(n_j-1)!}{(n_{j,q}-1)!\prod_{i=1\mbox{ and }i\neq q}^{|S|}n_{j,i}!}\\
                 & = & \sum_{q=2}^{|S|} \frac{(n_j-1)!n_{j,q}}{\prod_{i=1}^{|S|}n_{j,i}!}\\
                 & = &  \frac{(n_j-1)!}{\prod_{i=1}^{|S|}n_{j,i}!} \sum_{q=2}^{|S|} n_{j,q}
         \end{eqnarray*}

Consider $\overline{W}_{link}(o, s_1, k)$
\begin{eqnarray*}
  & =  & \prod_{j=1}^{k} \theta(o, s_1, j) \\
  & = & \prod_{j=1}^{k} [\frac{(n_j-1)!}{\prod_{i=1}^{|S|}n_{j,i}!} \sum_{q=2}^{|S|} n_{j,q}]\\
  & = & (\prod_{j=1}^{k} \frac{(n_j-1)!}{\prod_{i=1}^{|S|}n_{j,i}!}) (\prod_{j=1}^{k} \sum_{q=2}^{|S|} n_{j,q})\\
  & = & (\prod_{j=1}^{k} \frac{(n_j-1)!}{\prod_{i=1}^{|S|}n_{j,i}!}) (\prod_{j=1}^{k} (n_j - n_{j, 1}))
\end{eqnarray*}

From Equation~(\ref{eqn:breachProb}),
\begin{eqnarray*}
  &  & p(o,s_1,k)\\
 & = &  \frac{W_{link}(o, s_1, k)}{W_{total,k}} \\
          & = & \frac{W_{total, k} -  \overline{W}_{link}(o, s_1, k)}{W_{total,k}} \\
          & = & \frac{\prod_{j=1}^{k} \frac{n_j!}{\prod_{i=1}^{|S|}n_{j,i}!} - (\prod_{j=1}^{k} \frac{(n_j-1)!}{\prod_{i=1}^{|S|}n_{ji}!}) (\prod_{j=1}^{k} (n_j - n_{j, 1}))}{\prod_{j=1}^{k} \frac{n_j!}{\prod_{i=1}^{|S|}n_{j,i}!}}\\
          & = & \frac{\prod_{j=1}^{k} n_j - \prod_{j=1}^{k} (n_j - n_{j,1})}{\prod_{j=1}^{k} n_j}
\end{eqnarray*}

\begin{lemma}[Closed Form of $p(o,s_1,k)$]
\begin{eqnarray}
p(o,s_1,k) = \frac{\prod_{j=1}^{k} n_j - \prod_{j=1}^{k} (n_j - n_{j,1})}{\prod_{j=1}^{k} n_j}
\label{eqn:simplifiedBreachProb}
\end{eqnarray}
\label{lemma:probWithFunctionF}
\end{lemma}

From Equation~(\ref{eqn:breachProb}), $p(o,s_1,k)$ is
defined with a conceptual terms with the total number of
possible worlds. Lemma~\ref{lemma:probWithFunctionF} gives
a closed form of $p(o,s_1,k)$. Given the information
of $n_j$ (i.e., the size of the anonymized group
in the $j$-th table) and $n_{j,1}$ (i.e., the number
of tuples in the anonymized group with sensitive value
$s_1$ in the $j$-th table), we can calculate 
$p(o,s_1,k)$ with its 
closed form directly.

\begin{example}[Two-Table Illustration] \em
Consider that we want to protect the linkage between an individual
and a sensitive value $s_1$. Suppose
$o$ appears in both published tables $T_1^*$ and $T_2^*$.
Let $\mathcal{AG}_1(o)$ and $\mathcal{AG}_2(o)$
be the anonymized groups in $T_1^*$ and $T_2^*$ containing
$o$.
Suppose both $\mathcal{AG}_1(o)$ and $\mathcal{AG}_2(o)$
are linked to $s_1$.

By the notation adopted in this paper, $n_k$ is the size of
$\mathcal{AG}_k(o)$ and $n_{k,1}$ is the total number of tuples in
$\mathcal{AG}_k(o)$ with sensitive value $s_1$.

By Lemma~\ref{lemma:probWithFunctionF}, we have
\begin{eqnarray*}
p(o,s_1,k)  & = &  \frac{n_1 n_2 - (n_1 - n_{1, 1})(n_2 - n_{2, 1}) }{n_1 n_2}\\
  &  = & \frac{n_{2,1}n_1 + n_{1,1}n_2 - n_{1,1}n_{2,1}}{n_1 n_2}
\end{eqnarray*}
\done
\label{example:twoTable}
\end{example}

\begin{example}[Running Example] \em
In our running example as shown in Figure~\ref{tab:introExample2},
consider the second individual $o_2$ and a sensitive value
``chlamydia". We know that $n_1 = n_2 = 2$. Suppose $s_1$ is
``chlamydia". Thus, $n_{1,1}  = n_{2,1}
 = 1$. With respect to the published tables as shown in
Figure~\ref{tab:introExample2},
according to the formula derived in Example~\ref{example:twoTable},
$$
p(o_2,s_1,2) = \frac{1 \times 2 + 1 \times 2 - 1\times 1}{2 \times
2} = \frac{3}{4}
$$
which is greater than $1/2$ (the desired threshold).

However, if we publish tables as shown in Figure~\ref{size4}, then
$n_1 = n_2 = 4$ and $n_{1,1} = n_{2,1}  = 1$.
$$
p(o_2,s_1,2) = \frac{1 \times 4 + 1 \times 4 - 1\times 1}{4 \times
4} = \frac{7}{16}
$$
which is smaller than $1/2$.

In this example, we observe that, since
the published tables as shown in Figure~\ref{size4}
have a larger anonymized group size
(compared with
the published tables as shown in
Figure~\ref{tab:introExample2}), $p(o_2,s_1,2)$
is smaller.
\label{ex:runningExampleFormula}
\done

In this paper, we aim to publish table $T_k^*$ like
Figure~\ref{size4} at each time point $k$ such that $p(o,s,k) \le
1/\ell$ for each individual $o$ and each sensitive value $s$.
\label{example:probRunningExample}
\end{example}

From Example~\ref{ex:runningExampleFormula}, we observe
that a larger anonymized group size reduces the breach
probability that individual $o$ is linked to
sensitive value $s_1$ in the past. However, the anonymized group size
alone cannot reduce the breach probability. Consider
that an anonymized group in published table $T_k^*$ contains all sensitive values $s_1$,
instead of distinct sensitive values. Even though this anonymized
group is larger, if it still contains all sensitive values $s_1$,
it is easy to verify that an individual $o$
in this anonymized group must be linked to $s_1$ in this table $T_k^*$.

In fact, the breach probability is determined by the \emph{anonymized group size
ratio}. The anonymized group size ratio is equal to
the anonymized group size divided by the total number
of tuples in this anonymized group with sensitive value $s_1$.
In Example~\ref{ex:runningExampleFormula}, since all
sensitive values are distinct in an anonymized group (i.e.,
the total number
of tuples in this anonymized group with sensitive value $s_1$ is equal to 1),
the anonymized group size ratio is equal to the anonymized
group size.
In the next section, we will
show that the larger anonymized group size ratio can reduce the probability.

\section{Theoretical Properties}
\label{sec:properties}

In the previous section, we describe that a larger
anonymized group ratio can reduce the breach probability.
In this section, we will first study some properties of our problem,
including a \emph{minimum} anonymized group ratio for global privacy
guarantee, and then a monotonicity property that can be useful in data
anonymization.

\subsection{Minimum $\cal AG$ size Ratio}
\label{subsec:minEqClasSize}

Recall that $n_k$ is the anonymized group ($\cal AG$) size and
$n_{k,1}$ is the number of tuples in the anonymized group with
sensitive value $s_1$. In the following, we will derive the minimum
anonymized group size ratio $\frac{n_k}{n_{k,1}}$ for privacy
protection under the global guarantee.

\begin{theorem}
Let $k$ be an integer greater than 1. Suppose the anonymized group
in $T_k^*$ containing individual $o$ is linked to $s_1$. $p(o,s_1,k) \le
1/\ell$ if and only if
\begin{eqnarray}
\label{thm1} \frac{n_k}{n_{k,1}} \ge  \frac{\ell \prod_{j=1}^{k-1}(n_j - n_{j, 1})}{\ell \prod_{j=1}^{k-1}(n_j - n_{j, 1}) - (\ell -1) \prod_{j=1}^{k-1}n_j}
\end{eqnarray}
\label{thm:minSizeLinkWithS}
\end{theorem}
\textbf{Proof:} By Lemma~\ref{lemma:probWithFunctionF}, $p(o,s,k)$
is equal to $\frac{\prod_{j=1}^{k} n_j - \prod_{j=1}^{k} (n_j - n_{j,1})}{\prod_{j=1}^{k} n_j}$.
 $$p(o,s_1,k)  \le  1/\ell$$
 \begin{eqnarray*}
\Leftrightarrow  \frac{\prod_{j=1}^{k} n_j - \prod_{j=1}^{k} (n_j - n_{j,1})}{\prod_{j=1}^{k} n_j} & \le & \frac{1}{\ell} \\
\Leftrightarrow \frac{n_k \prod_{j=1}^{k-1}n_j - (n_k - n_{k,1})\prod_{j=1}^{k-1}(n_j - n_{j,1})}{n_k \prod_{j=1}^{k-1}n_j} & \le & \frac{1}{\ell} \\
\Leftrightarrow \frac{\prod_{j=1}^{k-1}n_j - (1 - \frac{n_{k,1}}{n_k})\prod_{j=1}^{k-1}(n_j - n_{j,1})}{ \prod_{j=1}^{k-1}n_j} & \le & \frac{1}{\ell}
\end{eqnarray*}
\begin{eqnarray*}
\Leftrightarrow \frac{n_{k, 1}}{n_{k}} & \le & 1 - \frac{\ell \prod_{j=1}^{k-1}n_j - \prod_{j=1}^{k-1}n_j}{\ell \prod_{j=1}^{k-1}(n_j - n_{j,1})}\\
\Leftrightarrow \frac{n_k}{n_{k,1}} & \ge & \frac{\ell \prod_{j=1}^{k-1}(n_j - n_{j, 1})}{\ell \prod_{j=1}^{k-1}(n_j - n_{j, 1}) - (\ell -1) \prod_{j=1}^{k-1}n_j}
\end{eqnarray*}
\done


From the above, for any $k > 1$, we can see that the value of $\frac{n_k}{n_{k,1}}$
should be lower bounded by the value of
$$\widetilde{\underline{n}}(k)  = \frac{\ell \prod_{j=1}^{k-1}(n_j - n_{j, 1})}{\ell \prod_{j=1}^{k-1}(n_j - n_{j, 1}) - (\ell -1) \prod_{j=1}^{k-1}n_j}$$
%
We define $ \widetilde{\underline{n}}(k) = \ell$ when $k = 1$.

\begin{example}[Running Example] \em
From Example~\ref{example:probRunningExample}, we know that the
published tables shown in Figure~\ref{size4} satisfy the privacy
requirement (i.e., $p(o,s,k) \le 1/\ell$ where $k=2$ and $\ell=2$).
At time $k=3$, we want to publish a new table $T_3^*$ from a raw
table $T_3$ which contain $o_2$.

Suppose we will put $o_2$ in the anonymized group
$\mathcal{AG}_3(o_2)$ in $T_3^*$ which is linked to $s_1$ where
$s_1$ is chlamydia.  By
Theorem~\ref{thm:minSizeLinkWithS}, when $k=3$, the R.H.S. of
Equation~(\ref{thm1}) becomes
\begin{eqnarray*}
 &  &  \frac{\ell (n_1 - n_{1, 1})(n_2 - n_{2, 1})}{\ell (n_1 - n_{1, 1})(n_2 - n_{2, 1}) - (\ell -1)n_1 n_2}\\
 & = & \frac{2(4-1)(4-1)}{2(4-1)(4-1) - (2-1)\cdot 4 \cdot 4}\\
 & = & 9
\end{eqnarray*}
which is the minimum anonymized group size ratio
$\frac{n_3}{n_{3,1}}$ in the published table $T_3^*$. Suppose
$\mathcal{AG}_3(o)$ contains only one occurrence of $s_1$. Then, the
size of the anonymized group $\mathcal{AG}_3(o)$ should be at least
$9$ so that $p(o, s_1, 3) \le 1/2$. \done
\end{example}

We have the following corollary when the inequality in
Theorem~\ref{thm:minSizeLinkWithS} becomes an equality.

\begin{corollary}
$\frac{n_k}{n_{k,1}} = \widetilde{\underline{n}}(k)$ if and only if
$p(o,s_1,k) = 1/\ell$.
\label{corollary:minSizeThenEqualToThreshold-new}
\end{corollary}

When a record for individual $o$ appears in a data release $T_i$ and
in the published data $T_i^*$, the anonymized group containing $o$ has no relation to
sensitive value $s$, then intuitively, this release should
not have any impact on the privacy protection of $o$ linking to $s$.
This is formally stated in the following lemma.

\begin{lemma}
If the anonymized group in $T_k^*$ containing $o$ is not linked to
$s_1$, then
$p(o,s_1,k) = p(o,s_1,k-1)$. 
\label{thm:minSizeNotLinkWithS}
\end{lemma}
\textbf{Proof:}
Since the anonymized group in $T_k^*$ containing $o$
is not linked to $s_1$, we know that $o$ is linked to $s_1$
in one of the first $(k-1)$-th published tables. Thus,
$$
     W_{link}(o, s_1, k) = W_k \times W_{link}(o, s_1, k-1)
$$
Thus, we have
\begin{eqnarray*}
p(o,s,k) & = & \frac{W_{link}(o, s_1, k)}{W_{total,k}} \\
   & = & \frac{W_k \times W_{link}(o, s_1, k-1)}{ \prod_{j=1}^k W_j }\hspace*{0.5cm}\mbox{(From Equation~(\ref{eqnarray:totalNoOfWorldMultipleTable}))}\\
   & = & \frac{W_{link}(o, s_1, k-1)}{W_{total,k-1}}\\
& = & p(o,s,k-1)
\end{eqnarray*}
\done


Thus, $\frac{n_k}{n_{k,1}}$ can be equal to any real number and does
not affect the value of $p(o,s,k)$ in this case.

Suppose a published table $T_k^*$ contains $o$ and we need to
generate an anonymized group $G$ containing $o$. Note that the size
of the anonymized group $G$ is $n_k$ and the number of tuples in $G$
with sensitive value $s_i$ is equal to $n_{k,i}$ for $i = [1, |S|]$.
Without loss of generality, suppose we want to protect the privacy linkage between an individual
$o$ and a sensitive value $s_1$.
From Theorem~\ref{thm:minSizeLinkWithS} and
Lemma~\ref{thm:minSizeNotLinkWithS}, we can determine the minimum
value of $\frac{n_k}{n_{k,1}}$ for generating an anonymized group
$G$. From Theorem~\ref{thm:minSizeLinkWithS}, if $G$ contains $s_1$,
in order to guarantee $p(o,s_1,k) \le 1/\ell$, we have to set the
value of $n_k$ to satisfy
$$
\frac{n_k}{n_{k,1}} \ge  \widetilde{\underline{n}}(k)
$$
From Lemma~\ref{thm:minSizeNotLinkWithS}, if $G$ does not contain
$s_1$, any value of $\frac{n_k}{n_{k,1}}$ will not affect the
privacy related to $o$ and $s_1$.

Although Theorem~\ref{thm:minSizeLinkWithS} suggests that if we set
the value of $\frac{n_k}{n_{k,1}}$ at least
$\widetilde{\underline{n}}(k)$, then $p(o,s_1,k) \le 1/\ell$.
However, suppose we set this value \emph{exactly} equal to
$\widetilde{\underline{n}}(k)$, although we can guarantee
$p(o,s_1,k) \le 1/\ell$ for the $k$ published tables, there will a
privacy breach (i.e., $p(o,s_1,k') > 1/\ell$) for any additional
\emph{future} published tables in which an anonymized group
containing $o$ is linked to $s_1$. This is a result of the following
lemma.

%

\begin{theorem}
Consider that we published $k-1$ tables where an anonymized group in
$T_{k-1}^*$ containing $o$ is linked to $s_1$. Suppose we are to
publish $T_k^*$ where an anonymized group in $T_k^*$ containing $o$
is also linked to $s_1$. If $\frac{n_{k-1}}{n_{k-1,1}} =
\widetilde{\underline{n}}(k-1)$, then $p(o,s_1,k) > 1/\ell$.
\label{lemma:setMinimumSizePrivacyBreach-new}
\end{theorem}

\subsection{Monotonicity}
\label{subsec:monotonicity}

Monotonicity is a useful property for some anonymization process
where the resulting anonymization groups are constructed in a
bottom-up manner, merging smaller groups that violates the privacy
requirement into bigger groups which may guarantee privacy. It is
also useful when the anonymization is top-down, splitting bigger
groups into smaller ones as long as the privacy guarantee holds.

Consider the privacy protection for the linkage of an individual $o$
to a sensitive value $s_1$. From
Lemma~\ref{thm:minSizeNotLinkWithS}, we know that $p(o,s_1,k)$ is
\emph{independent} of data releases in which any anonymized group
containing $o$ (in the published tables) are not linked to $s_1$.
Hence, in the following, we consider the worst-case scenario where in
all releases whenever there exists an anonymized group containing $o$
(in a published table), $o$ is linked to $s_1$.

The monotonicity property is described as follows.

\begin{theorem}[Monotonicity]
$p(o,s_1,k)$ is strictly decreasing when $\frac{n_k}{n_{k,1}}$
increases. \label{lemma:monotonicity}
\end{theorem}

The proof is given in the appendix. Note that $n_k/n_{k,1}$ is
essentially the inverse of the proportion of $s_1$ tuples in the
anonymized group. Therefore, when a bigger group that satisfies the privacy
requirement is
split into smaller ones, if the proportion of $s_1$ tuples in the small group containing $o$
is not increased, then $p(o,s_1,k)$ is not increased.
Conversely if a small group violates the privacy guarantee, merging
it with another group may decrease the proportion of $s_1$ tuples and thus $p(o,s,k)$ may be decreased.

An anonymized group $\mathcal{AG}$ is said to violate
the global guarantee if there exists an individual $o \in \mathcal{AG}.I$
and a sensitive value $s \in \mathcal{AG}.S$ such that
$p(o,s,k) > 1/\ell$.

\begin{corollary}
Consider an anonymized group $\mathcal{AG}$ in 
the published table $T_k^*$ which violates
the global guarantee. 
If we partition $\mathcal{AG}$ into a number
of smaller groups,
one of the smaller groups violates the global guarantee. 
\end{corollary}
\textbf{Proof Sketch:} Suppose $n_k/n_{k,1}$ is the size ratio
for $\mathcal{AG}$. It is easy to see that
one of the smaller groups has the size ratio smaller
than $n_k/n_{k,1}$. By Theorem~\ref{lemma:monotonicity},
$p(o, s, k)$ increases. Since 
$\mathcal{AG}$ violates the global guarantee (i.e., 
$p(o,s,k) > 1/\ell$), the smaller group also violates
the global guarantee (i.e., $p(o,s,k) > 1/\ell$). 
\done

\section{Anonymization}
\label{sec:alg}

In previous sections, we have observed that, by choosing a proper
size of an anonymized group, the global privacy guarantee can be
achieved. In general, a size above a certain threshold size can be
chosen. However, setting a size equal to the threshold size will
make future anonymization infeasible (see
Theorem~\ref{lemma:setMinimumSizePrivacyBreach-new}). Therefore, it
is necessary to choose a size that is greater than the threshold.
The increase in size however, would lead to a decrease in the
utility of the data. Hence, a question will be how to pick a smallest
size that can maintain the global guarantee.

In this section, we show that if we are given a bound on the number
of releases where an individual $o$ may be linked to a sensitive
value $s$, then we can devise a strategy to minimize the maximum
anonymization group size. We also propose another strategy which
aims to reduce the anonymized group size on average.


%

\subsection{Constant-Ratio Strategy} \label{subsec:equalSizeStrategy}

In database related problems, one can typically derive effective
mechanisms based on the characteristics of the data itself. In our
problem scenario, a data publisher has at his/her disposal the
statistical information of the data collections. For example,
consider the medical database. The statistics can point to the
expected frequency of an individual contracting a certain disease
over his or her lifespan. With such information, one can set an
estimated bound on the number of data releases that a person may
indeed be linked to the
disease. 
With this knowledge, one can adopt a constant-ratio strategy which we
shall show readily can minimize the maximum size of the
corresponding anonymized groups.

\emph{Constant-ratio strategy} makes sure that the size of
anonymized groups ${\cal AG}(o)$ for individual $o$ containing $s_1$
divided by the number of occurrences of $s_1$ remain unchanged over
a number of data releases. Formally, given an integer $k'$ for the number of
data releases, 
for $i \in [1, k']$,
$$
\frac{n_{o_i}}{n_{o_i,1}} = \widetilde{n}_c
$$
where $\widetilde{n}_c$ is a positive real number constant, and
$o_i$ is a timestamp for the $i$-th release where both $o$ and $s_1$
appear. For the sake of simplicity, we set
$
  \widetilde{n}_c = \frac{n}{n_s}
$ where $n$ and $n_s$ are positive integer constants where $n_s \le
n$.

$k'$ corresponds to the total number of possible releases in the future.
In other words, during data publishing,
the data publisher expects to publish $k'$
table for this data. With this given parameter $k'$,
we can calculate $n$ and $n_s$ such that
$\frac{n_{o_i}}{n_{o_i,1}}$ remain unchanged when $i$ changes.

In order to make sure that $p(o,s_1,j) \le 1/\ell$ for any $j \in
[1, k']$, we need to protect $p(o,s_1,k') \le 1/\ell$.
In the following, we consider $p(o,s_1,k')$ which is equal to
\begin{eqnarray*}
\frac{\prod_{j=1}^{k'} n_j - \prod_{j=1}^{k'} (n_j - n_{j,1})}{\prod_{j=1}^{k'} n_j} & \le & \frac{1}{\ell} \\
n^{k'} - (n - n_s)^{k'} & \le & n^{k'} \times \frac{1}{\ell} \\
1 - (1 - \frac{n_s}{n})^{k'} & \le & \frac{1}{\ell} \\
\frac{n_s}{n} & \le & 1 - (1 - \frac{1}{\ell})^{1/k'} \\
\frac{n}{n_s} & \ge & 1/[1 - (1 - \frac{1}{\ell})^{1/k'}]
\end{eqnarray*}

Let $\widetilde{n}_c = 1/[1 - (1 - \frac{1}{\ell})^{1/k'}]$.
%
%
%
%
%
%

Table~\ref{tab:valueForConstantStrategy} shows
the values of $\widetilde{n}_c$ with selected values of $\ell$ and $k'$.
When $\ell$ increases, $\widetilde{n}_c$ increases.
When $k'$ increases, $\widetilde{n}_c$ also increases.

\begin{table}
\center\scriptsize
\begin{tabular}{| c ||r|r|r|r|r|r|r|r|}\hline
  $\ell$ & 2 & 2 & 2&5&10&2&5&10\\ \hline
  $k'$ & 2 & 5 & 20 &20&20&10&10&10 \\ \hline
  $\widetilde{n}_c$ & 3.44 & 7.75 & 29.41 & 90.13 & 190.33 & 15.10 &
  45.35 & 95.42 \\ \hline
\end{tabular}
\caption{Values of $\widetilde{n}_c$ with selected values of $l$ and
$k'$} \label{tab:valueForConstantStrategy}
\end{table}


It remains to show that the constant-ratio strategy indeed can lead
to data publishing that minimizes the maximum anonymized group sizes.
First, we define this property more formally.

\begin{definition}[Min-Max optimization] An anonymization for serial
data publishing is min-max optimal if the maximum anonymized group
size among the anonymized groups containing individual $o$ and
sensitive value $s_1$ for any given $o$ and $s_1$ over all data
releases is minimized.
\end{definition}

\begin{theorem}[Optimality]
The constant-ratio strategy generates a min-max optimal solution for
serial data publishing. \label{lemma:equalSizeOptimal}
\end{theorem}
\textbf{Proof:} Let $N$ be the set of anonymized group sizes in the
$k'$ published tables where these anonymized groupes contain $o$ and
are linked to $s_1$. That is, $N = \{n_1, n_2, ..., n_{k'}\}$. Let
$u(N) = \max_{n_i \in N} n_i$. Let $N_a$ be the set of anonymized
group sizes in the $k'$ published tables generated by strategy $a$.

Let $p(o,s_1,k'|a)$ be $p(o,s_1,k')$ with respect to strategy $a$. Let $A$
be the set of all possible strategies $a$ such that, with the
published tables with strategy $a$, $p(o,s_1,k'|a) \le 1/\ell$. Suppose
$a_o$ is the constant-ratio strategy. We will prove that this
strategy can obtain an optimal value of $u(N)$. That is,
$$u(N_{a_o}) = \min_{a \in A}\{u(N_a)\}$$

We prove by contradiction. Consider that the strategy $a_o$
generates $N_{a_o} = \{n_1, n_2, ..., n_{k'}\}$. By
Corollary~\ref{corollary:minSizeThenEqualToThreshold-new}, it is
easy to verify that $p(o,s_1,k'|a_o) = 1/\ell$.

Suppose there exists a strategy $a' \neq a_o$ which generates
$N_{a'} = \{n_1', n_2', ..., n_{k'}'\}$ such that $u(N_{a'}) <
u(N_{a_o})$ and $p(o,s_1,k'|a') \le 1/\ell$. We deduce that, for all
$i \in [1, k']$,
$$
  n_i' < n_i
$$
By Theorem~\ref{lemma:setMinimumSizePrivacyBreach-new}, we know that,
$p(o,s_1,k'|a') > p(o,s_1,k'|a_o)$ (which is equal to $1/\ell$). We
conclude that $p(o,s_1,k'|a') > 1/\ell$. Thus, privacy breach occurs,
which leads to a contradiction. \done

Although the constant-ratio strategy generates a min-max optimal
solution, 
the statistical information about the data should be known. 
For example, the constant-ratio strategy requires the priori knowledge
about $k'$ which is equal to the total number of possible releases
in the future. If such information is unavailable, we can use
the \emph{geometric strategy} proposed in the next subsection
where this strategy does not require the statistical
information. 

\subsection{Geometric Strategy} \label{subsec:geometricStrategy}

Other than minimizing the maximum anonymized group size, another
desirable utility criterion will be to minimize the average group
size. In order to achieve this goal, we examine the probability of
occurrences of anonymized groups for linking individuals $o$ to a
certain sensitive value $s$. From past data, there will be a
distribution of the total number of releases where any given
individual has contracted disease $s$. For example, if the maximum
of such value is 10, some individuals may be linked to $s$ 10 times
in total, but most individuals may be linked to $s$ less than 10
times in total. Typically, the number of individuals that are linked
to $s$ for at least $k$ releases will be greater than that for $k''$
releases where $k < k'' $. Therefore, when choosing the sizes of the
anonymized groups, it will reduce the average group size if we
choose smaller sizes for the earlier releases where $o$ is linked to
$s$ and bigger sizes for the later such releases. This is the
essence of our next proposed strategy, namely, the \emph{geometric
strategy}.

With the geometric strategy, the anonymized group size will be equal
to the minimum feasible value of $\widetilde{\underline{n}}(k)$
multiplied by a factor, $\alpha$, at any time point $k$. This will
be a growing value since the value of $\widetilde{\underline{n}}$
will grow with $k$. Note that $\alpha$ must be greater than 1 since
with $\alpha = 1$, the minimum feasible
$\widetilde{\underline{n}}(k)$ will be used, and from Theorem
\ref{lemma:setMinimumSizePrivacyBreach-new}, that will make future
selection of group size infeasible. The value of $\alpha$ can be
selected based on the estimated number of releases where an
individual will be linked to $s$ in total.

Thus, with this strategy, with $j \ge 1$, we set
$$
\frac{n_j}{n_{j,1}} = \alpha \cdot \widetilde{\underline{n}}(j)
$$


Figure~\ref{fig:effectOfAlphaAndL} shows how the values of
$\frac{n_k}{n_{k,1}}$ increases with $k$.
Figure~\ref{fig:effectOfAlphaAndL}(a) studies the effect of $\alpha$
(with $\ell$ set to 2) and Figure~\ref{fig:effectOfAlphaAndL}(b)
studies the effect of $\ell$ (with $\alpha$ set to 5). When $k$
increases, $\frac{n_k}{n_{k,1}}$ increases. When $\alpha$ is larger,
although the initial value of $\frac{n_k}{n_{k,1}}$ is larger, the
grow rate of $\frac{n_k}{n_{k,1}}$ is smaller. When $\ell$
increases, $\frac{n_k}{n_{k,1}}$ increases.
Figure~\ref{fig:effectOfAlphaAndL}(a) shows that $\alpha = 5$ and 10
are better choices than $\alpha=3$ since the increase in the ratio
is much slower. Note that these values are all pre-computable and it
is easy to choose a suitable parameter by examining the pre-computed
trends.

\begin{figure}[tbp] 
\begin{tabular}{c c c c}
    \begin{minipage}[htbp]{4.1cm}
        \includegraphics[width=4.1cm,height=3.0cm]{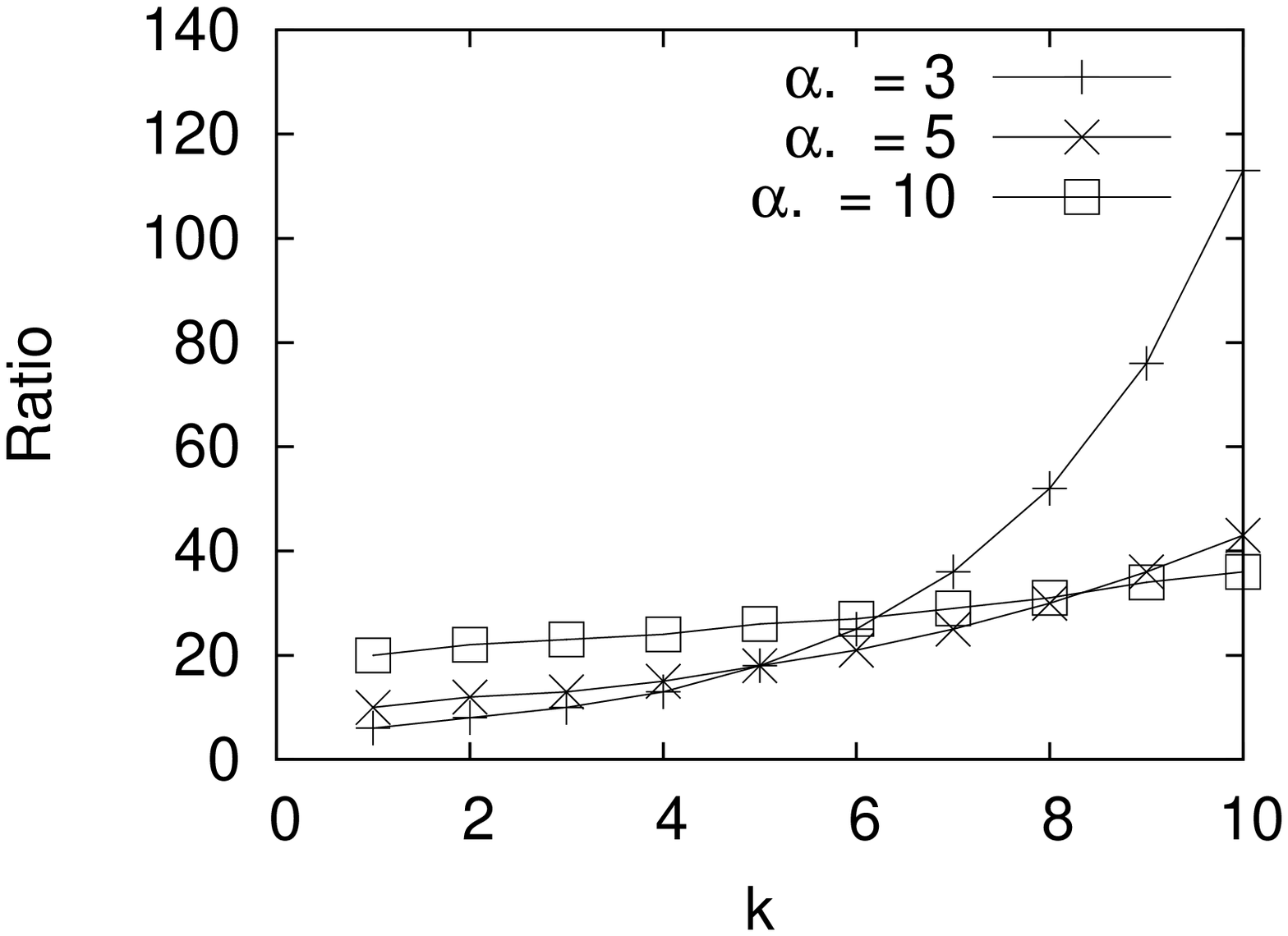}
    \end{minipage}
&
    \begin{minipage}[htbp]{4.1cm}
        \includegraphics[width=4.1cm,height=3.0cm]{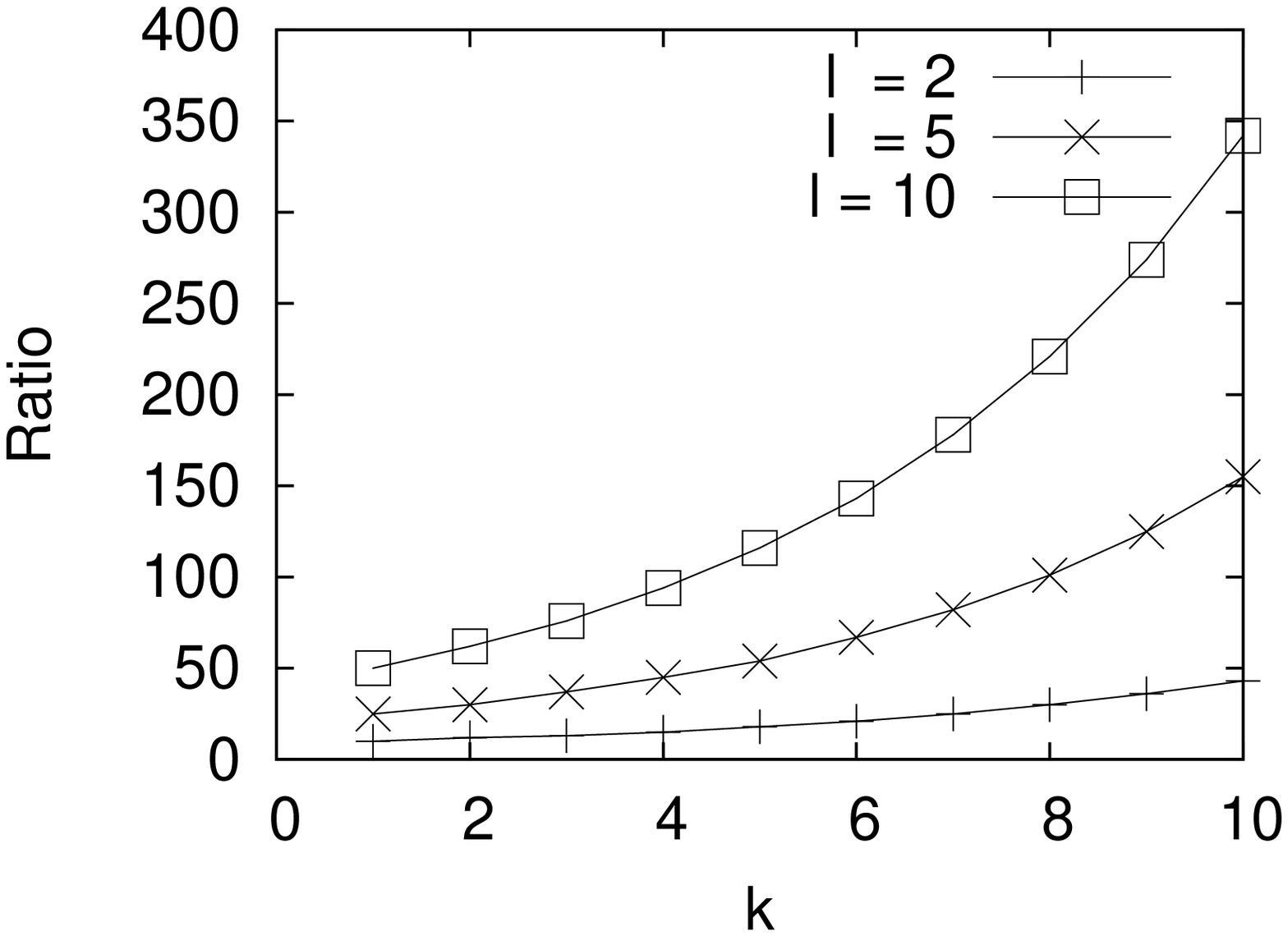}
    \end{minipage}
    \\
(a) &
 (b)
\end{tabular}
\caption{Effect of $\alpha$ and $l$ on ratio $n_k/n_{k,1}$}\label{fig:effectOfAlphaAndL}
\end{figure}

\begin{figure*}[htbp] 
\begin{tabular}{c c c c}
    \begin{minipage}[htbp]{4.1cm}
        \includegraphics[width=4.1cm,height=3.0cm]{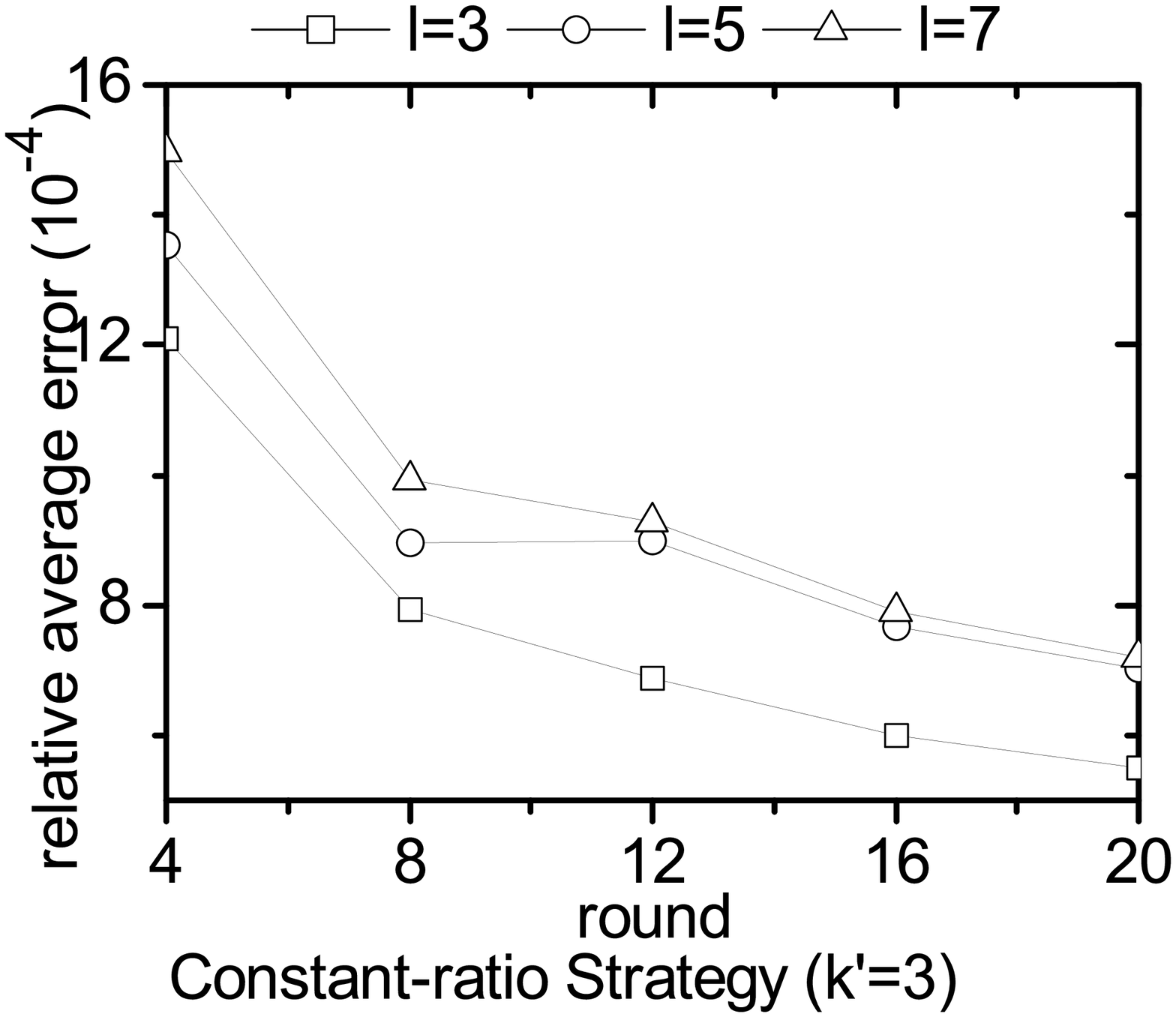}
    \end{minipage}
&
    \begin{minipage}[htbp]{4.1cm}
        \includegraphics[width=4.1cm,height=3.0cm]{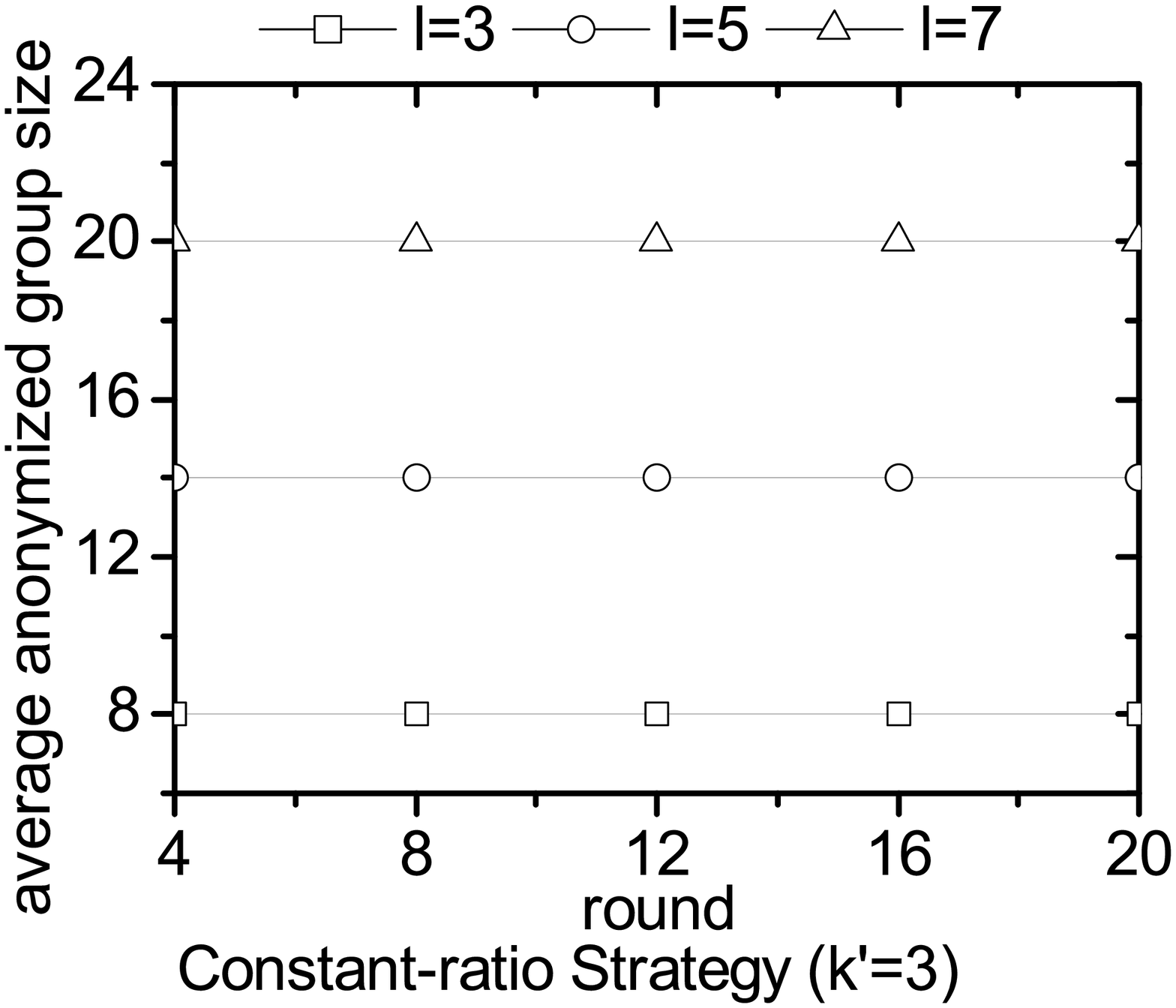}
    \end{minipage}
&
    \begin{minipage}[htbp]{4.1cm}
        \includegraphics[width=4.1cm,height=3.0cm]{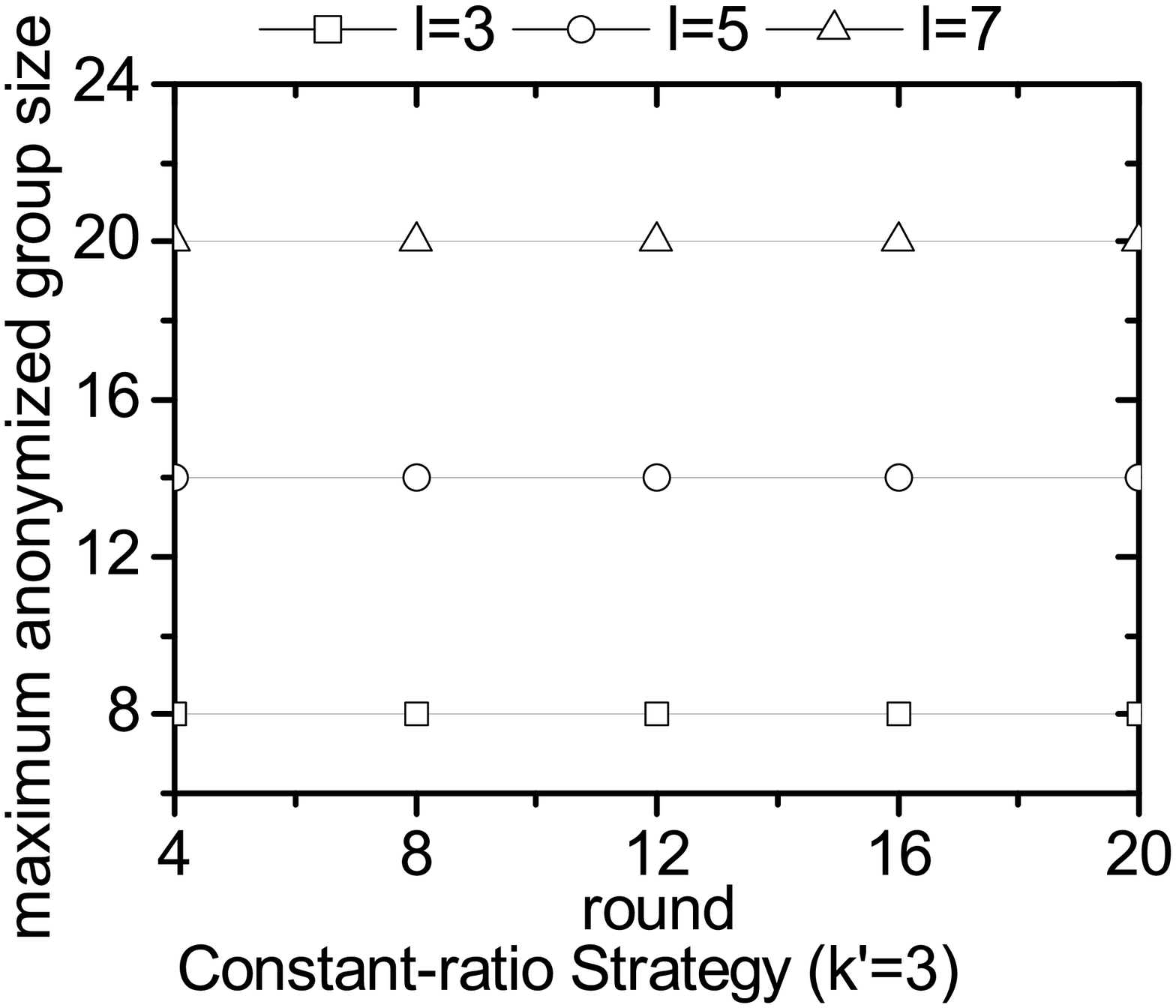}
    \end{minipage}
&
    \begin{minipage}[htbp]{4.1cm}
        \includegraphics[width=4.1cm,height=3.0cm]{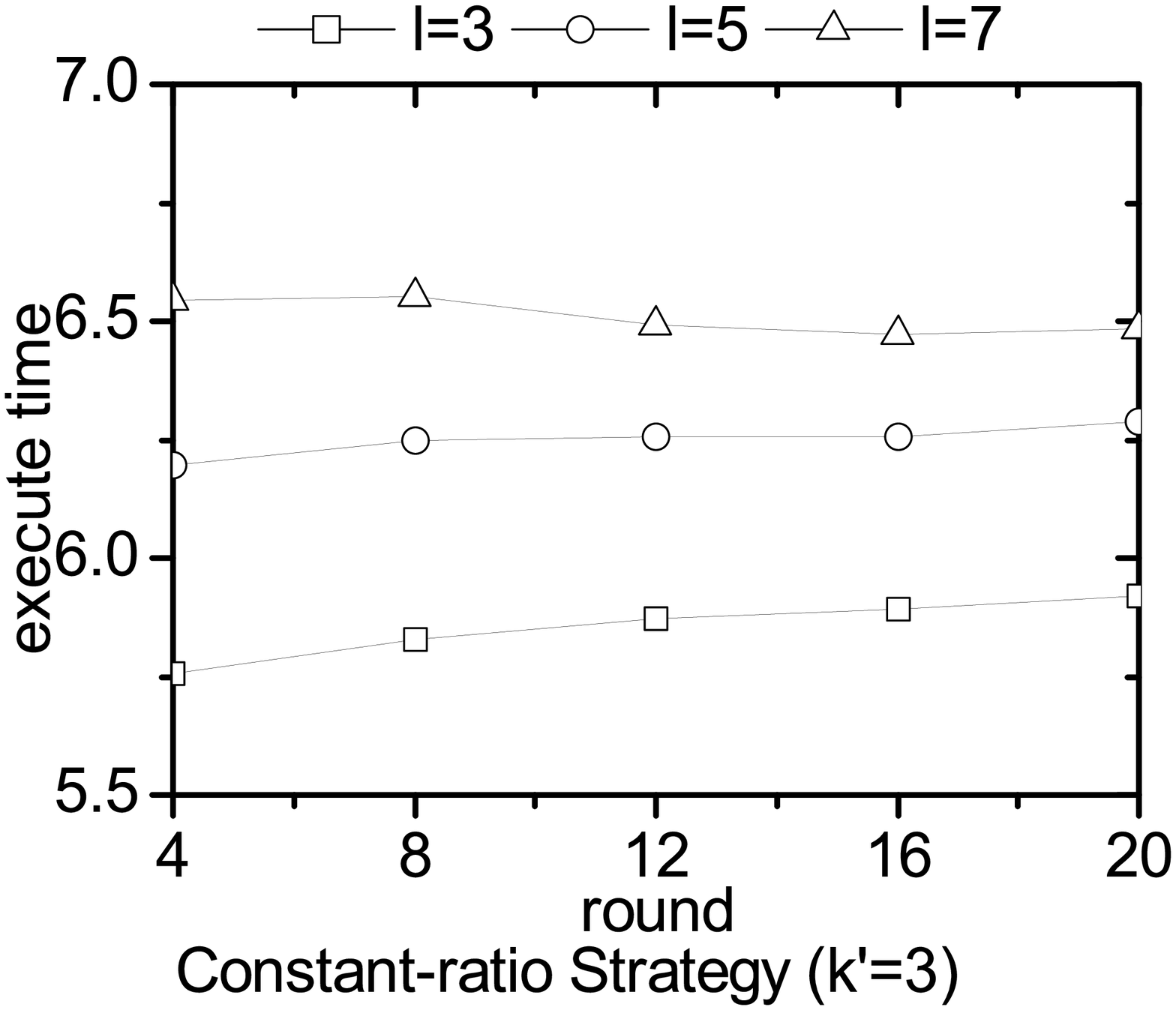}
    \end{minipage}
    \\
(a) &
 (b)
& (c) & (d)
\end{tabular}\vspace{-2mm}
\caption{Effect of $l$ (Constant-Ratio Strategy) where $k' =
3$}\vspace{-2mm}\label{fig:graphConstantForL}
\end{figure*}

\begin{figure*}[htbp] 
\begin{tabular}{c c c c}
    \begin{minipage}[htbp]{4.1cm}
        \includegraphics[width=4.1cm,height=3.0cm]{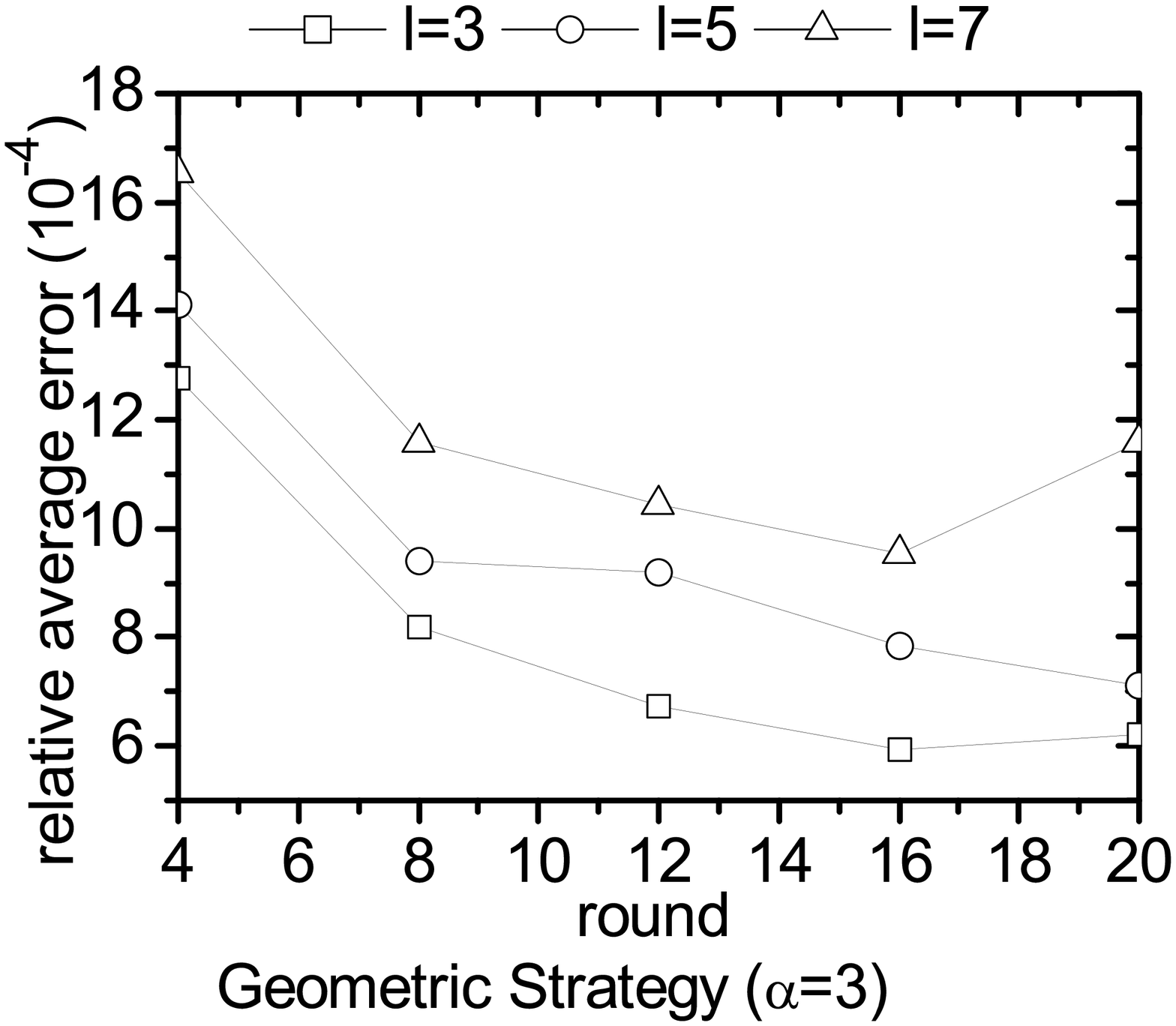}
    \end{minipage}
&
    \begin{minipage}[htbp]{4.1cm}
        \includegraphics[width=4.1cm,height=3.0cm]{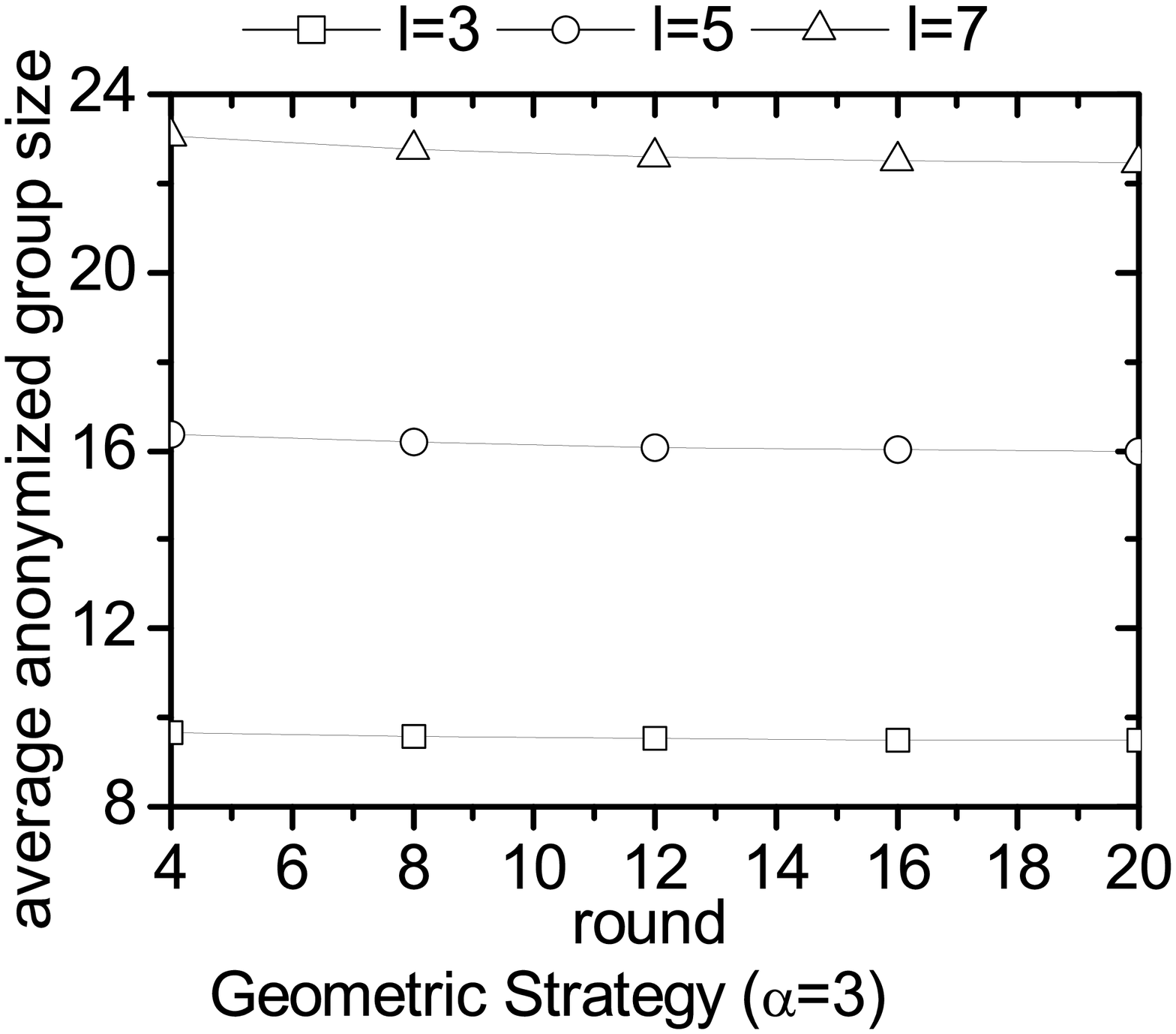}
    \end{minipage}
&
    \begin{minipage}[htbp]{4.1cm}
        \includegraphics[width=4.1cm,height=3.0cm]{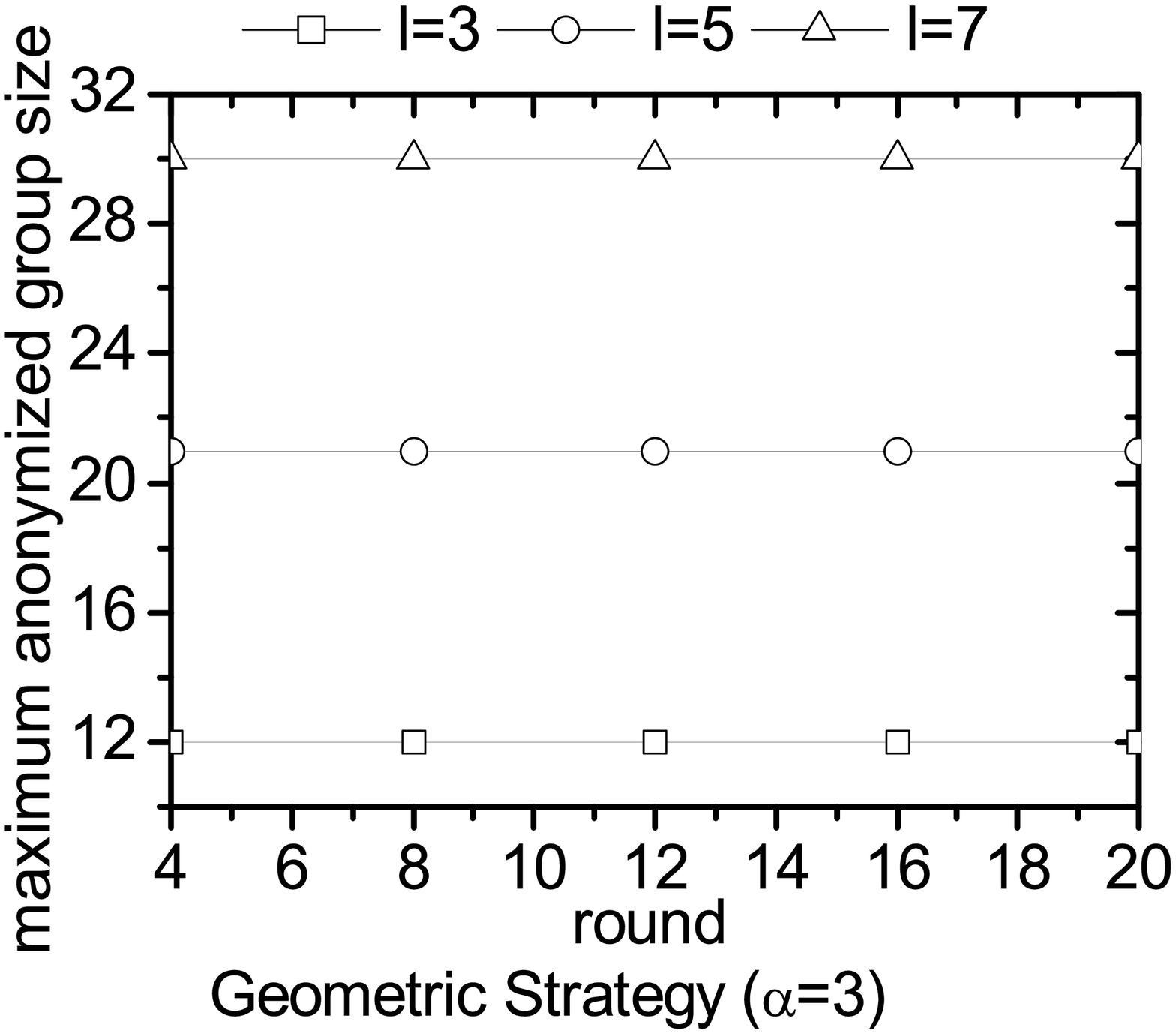}
    \end{minipage}
&
    \begin{minipage}[htbp]{4.1cm}
        \includegraphics[width=4.1cm,height=3.0cm]{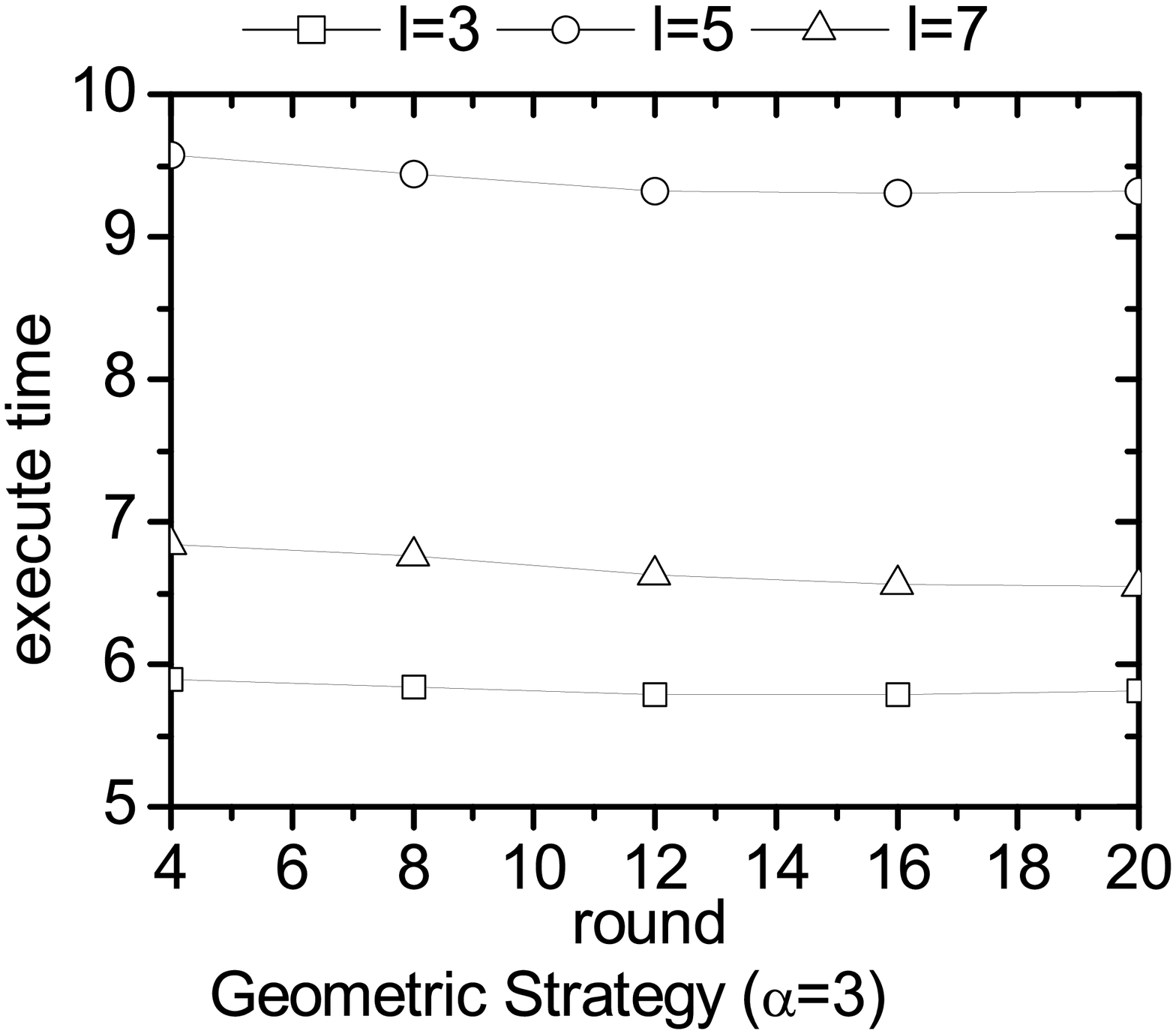}
    \end{minipage}
    \\
(a) &
 (b)
& (c) & (d)
\end{tabular}\vspace{-2mm}
\caption{Effect of $l$ (Geometric Strategy) where $\alpha =
3$}\vspace{-2mm}\label{fig:graphGeometricForL}
\end{figure*}

\subsection{Discussion}

In both of the above strategies, there may occur rare occasions
where the required anonymized group size is not available in the
given data set. As with previous works \cite{XT07,BFW+08}, we handle
the exceptional cases by data distortion. We can suppress the
sensitive values of individuals when it is found that no feasible
group size can maintain the global guarantee for privacy
preservation. From our experimental results, such suppression has
not been found needed.

Though our discussion has been based on a single value for the
sensitive attribute in each record, our results can be easily
extended to the case where each record may contain a \emph{set} of values
for the sensitive attribute. The essential proportion of possible
worlds where an individual is linked to a sensitive value would not
be affected.

\section{Implementation} \label{sec:implement}

In Section~\ref{sec:alg}, we describe two strategies to determine
the value of $\frac{n_k}{n_{k,1}}$ for privacy protection with
respect to a sensitive value $s_1$. In the following, we describe
how to anonymize the table given the desired value of
$\frac{n_k}{n_{k,1}}$.

Since the formula is based on the frequency that a tuple for
individual $o$ is linked to a sensitive value $s$ in an anonymized
group from published tables (by Theorem~\ref{thm:minSizeLinkWithS}
and Lemma~\ref{thm:minSizeNotLinkWithS}), we propose to keep a
data structure, called \emph{statistics file}, to store the sizes of
the anonymized groups containing a record for individual $o$ such
that
 $o$ is linked to a sensitive value $s$, denoted by $m(o, s)$.
Consider an individual $o$ and a sensitive value $s$. Let the
anonymized groups containing $o$ in $T_1^*, T_2^*$ and $T_3^*$ be
$G_1$ (of size 3), $G_2$ (of size 5) and $G_3$ (of size 4),
respectively. If $G_1$ and $G_2$ contain $s$ but $G_3$ does not,
$m(o, s)$ is equal to $\{3, 5\}$. Suppose there is another published
table $T_4^*$ which does not contain $o$. $m(o, s)$ is also equal to
$\{3, 5\}$.

Given the statistics file, it is possible to adopt existing known
anonymization methods to generate anonymized groups that satisfy the
group size ratio requirement of interest. For example, we may use a
bottom-up approach to grow the anonymized groups. Alternatively, we
can use a top-down approach to keep breaking up large anonymized
groups and stops when it begins to violate the group size ratio
requirement.

\section{Empirical Studies}
\label{sec:exp}

\begin{figure*}[htbp] 
\begin{tabular}{c c c c}
    \begin{minipage}[htbp]{4.2cm}
        \includegraphics[width=4.1cm,height=3.1cm]{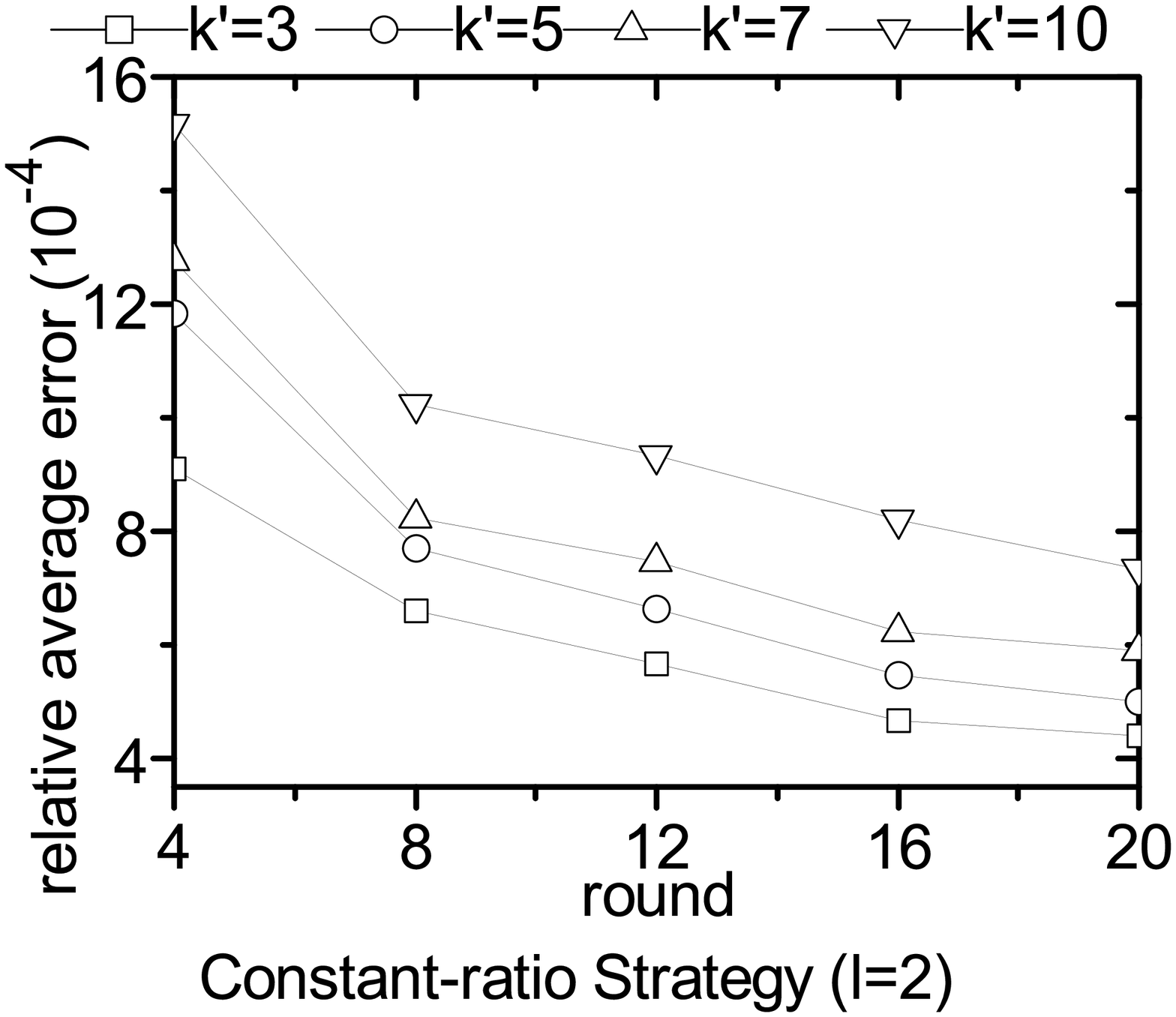}
    \end{minipage}
&
    \begin{minipage}[htbp]{4.2cm}
        \includegraphics[width=4.2cm,height=3.1cm]{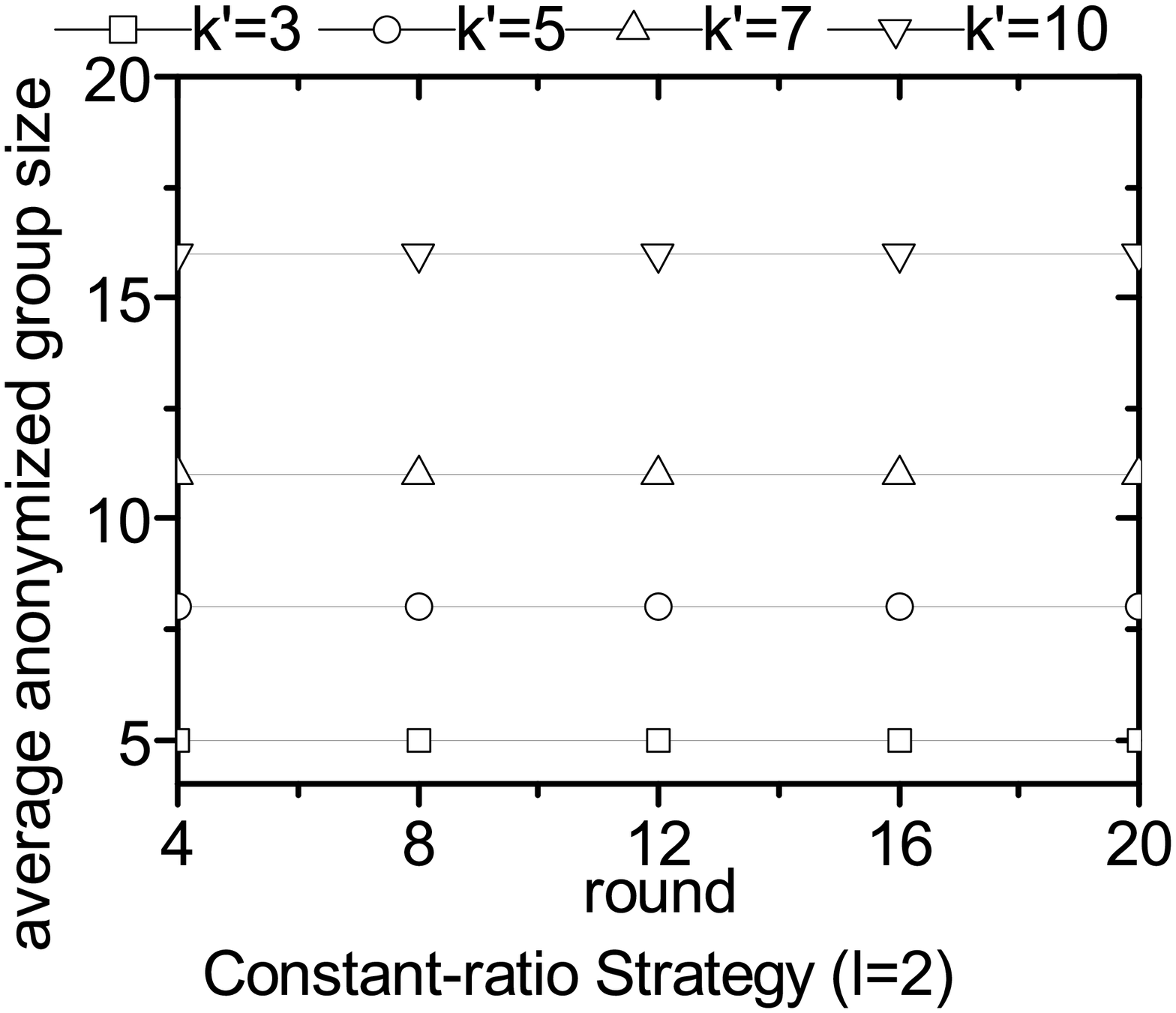}
    \end{minipage}
&
    \begin{minipage}[htbp]{4.2cm}
        \includegraphics[width=4.2cm,height=3.1cm]{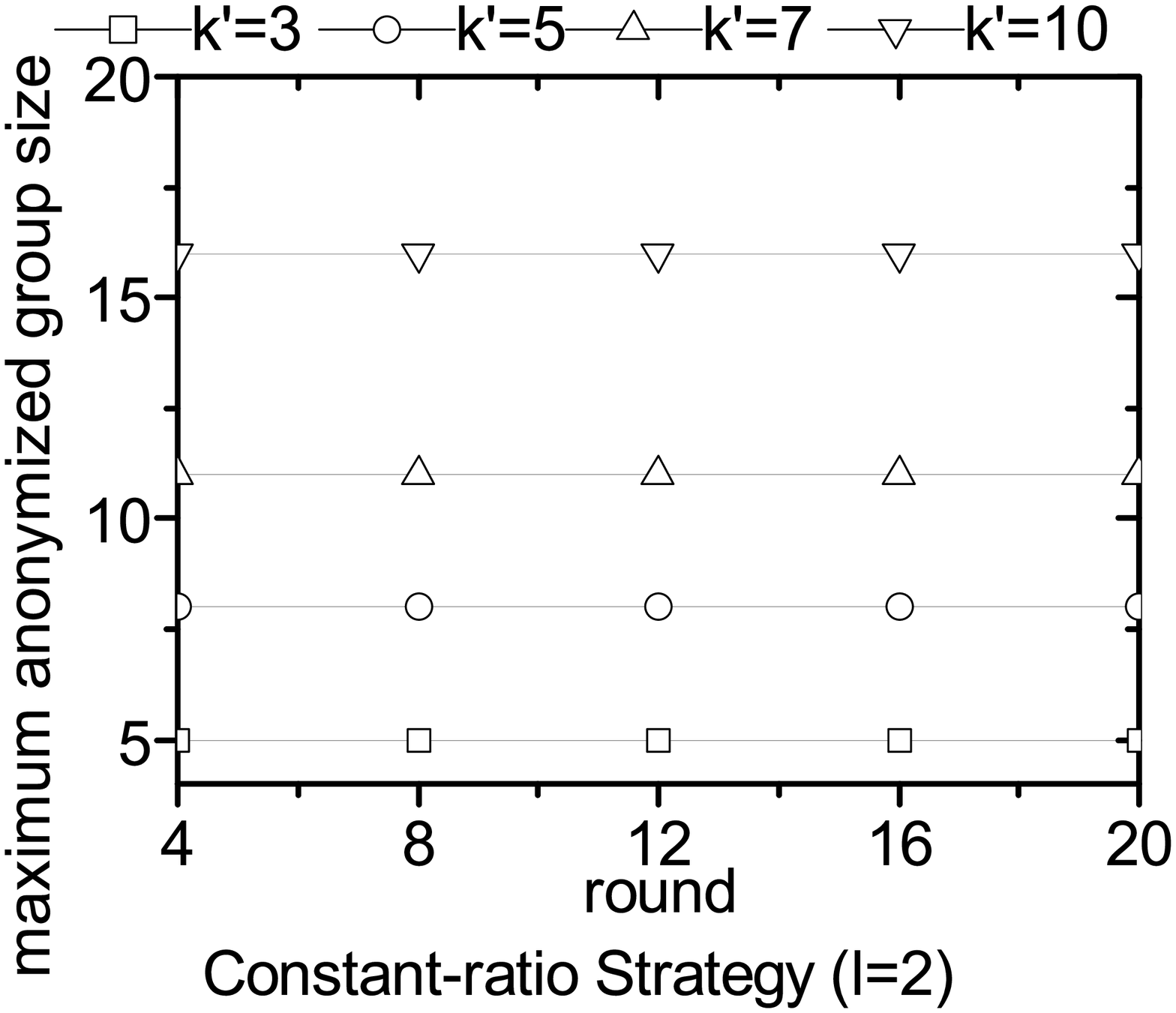}
    \end{minipage}
&
    \begin{minipage}[htbp]{4.2cm}
        \includegraphics[width=4.2cm,height=3.1cm]{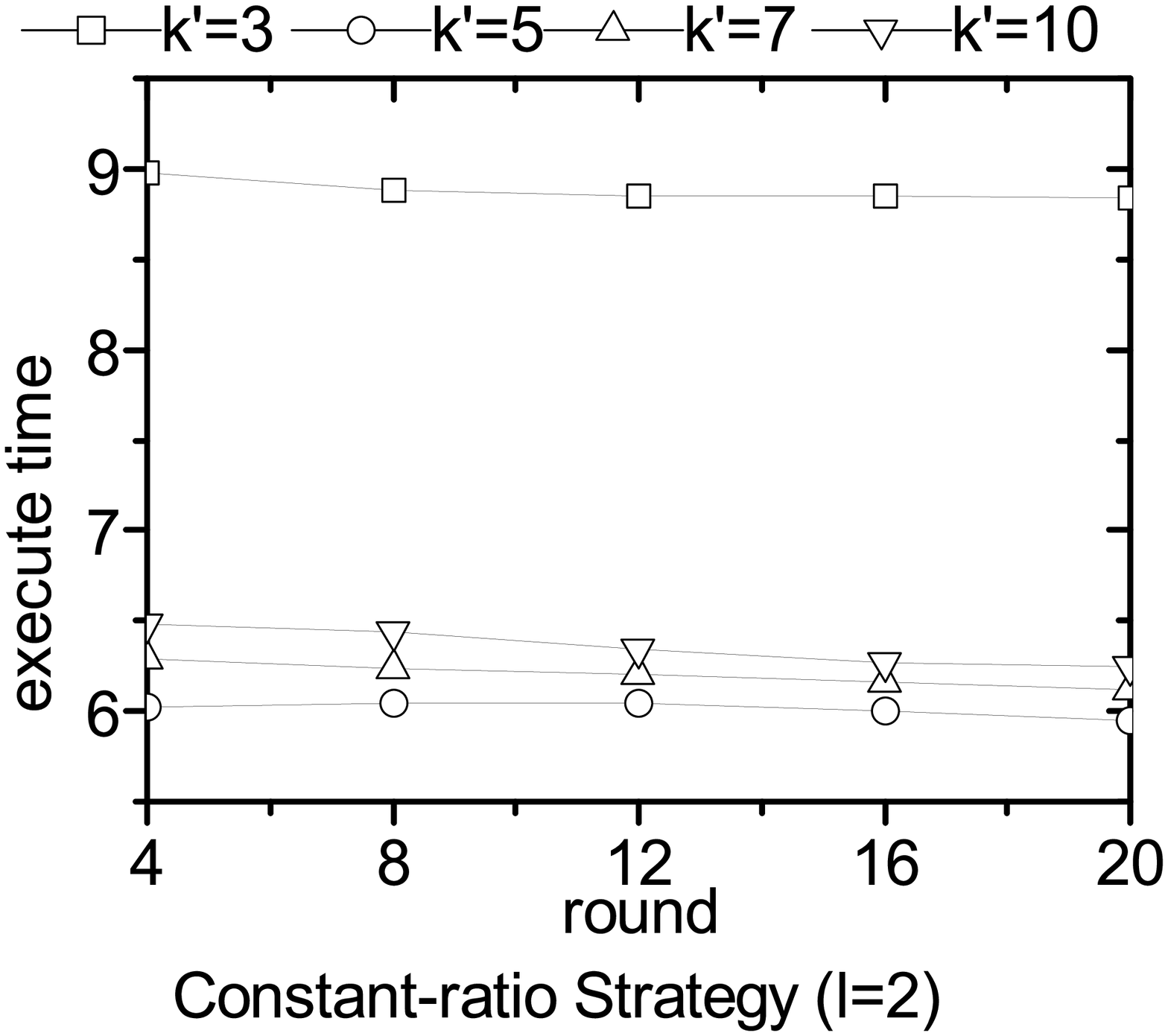}
    \end{minipage}
    \\
(a) &
 (b)
& (c) & (d)
\end{tabular}\vspace{-2mm}
\caption{Effect of $k'$ (Constant-Ratio Strategy) where $\ell =
2$}\vspace{-2mm}\label{fig:graphConstantForNewK}
\end{figure*}

\begin{figure*}[tbp] 
\begin{tabular}{c c c c}
    \begin{minipage}[htbp]{4.2cm}
        \includegraphics[width=4.2cm,height=3.1cm]{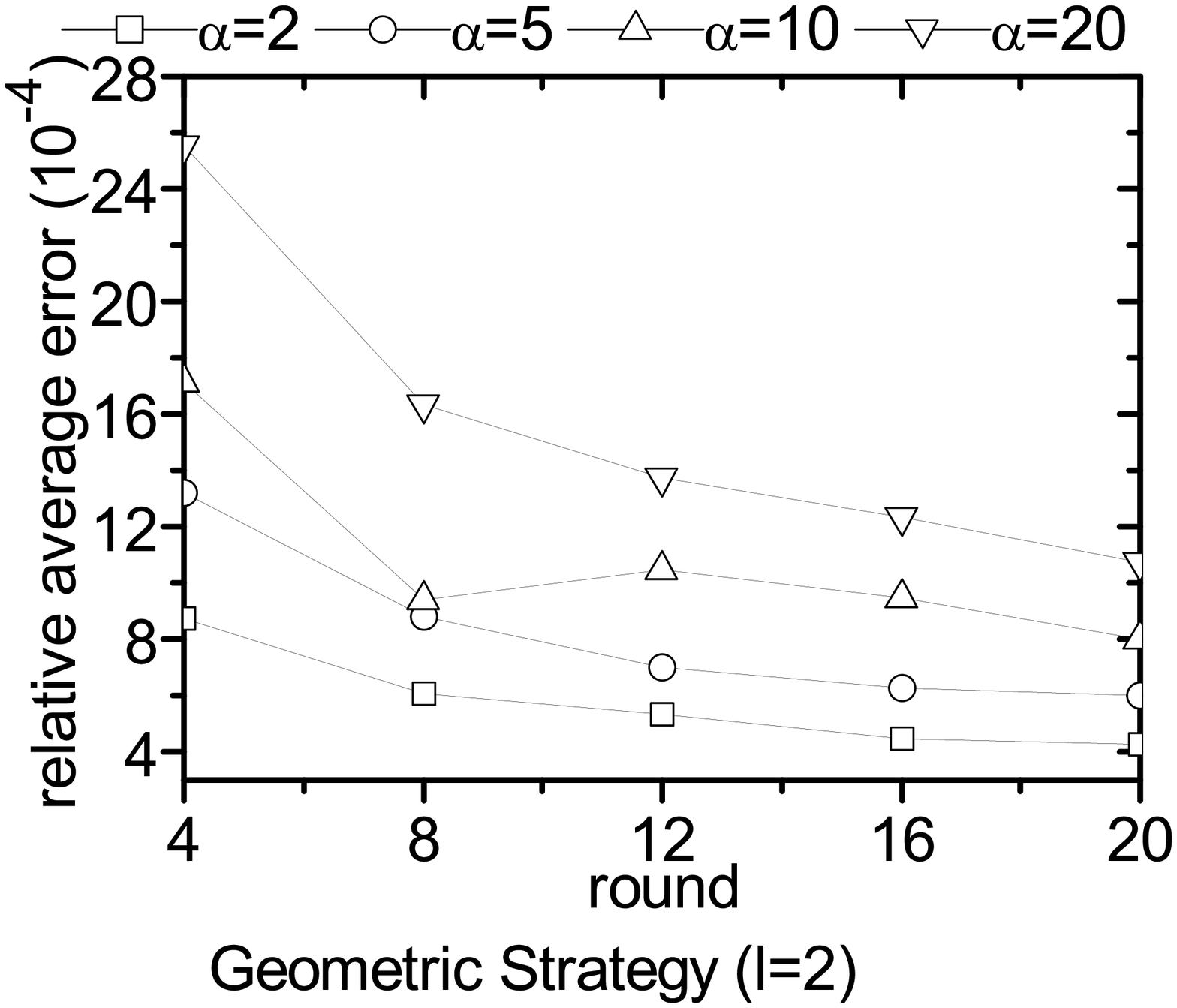}
    \end{minipage}
&
    \begin{minipage}[htbp]{4.2cm}
        \includegraphics[width=4.2cm,height=3.1cm]{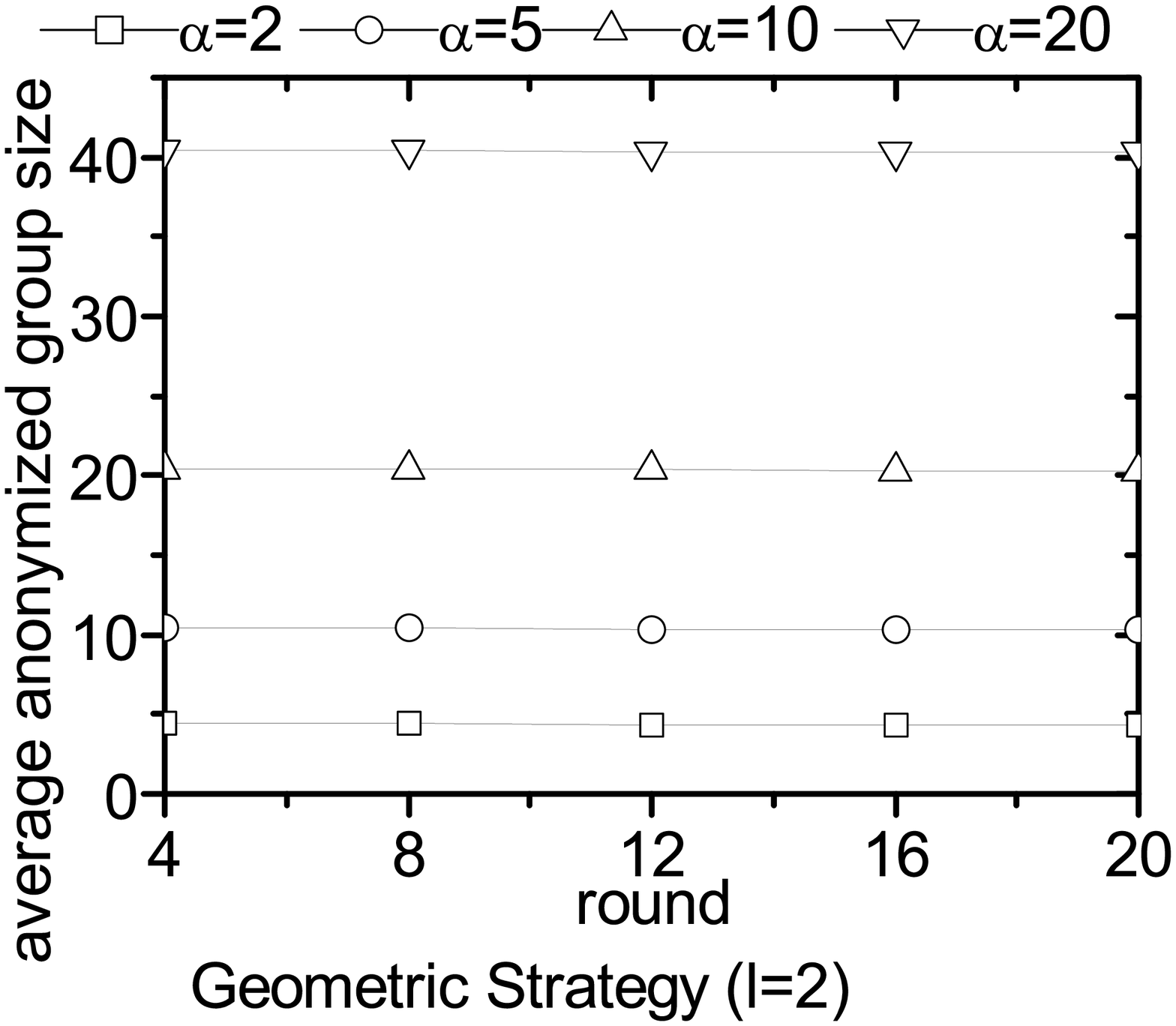}
    \end{minipage}
&
    \begin{minipage}[htbp]{4.2cm}
        \includegraphics[width=4.2cm,height=3.1cm]{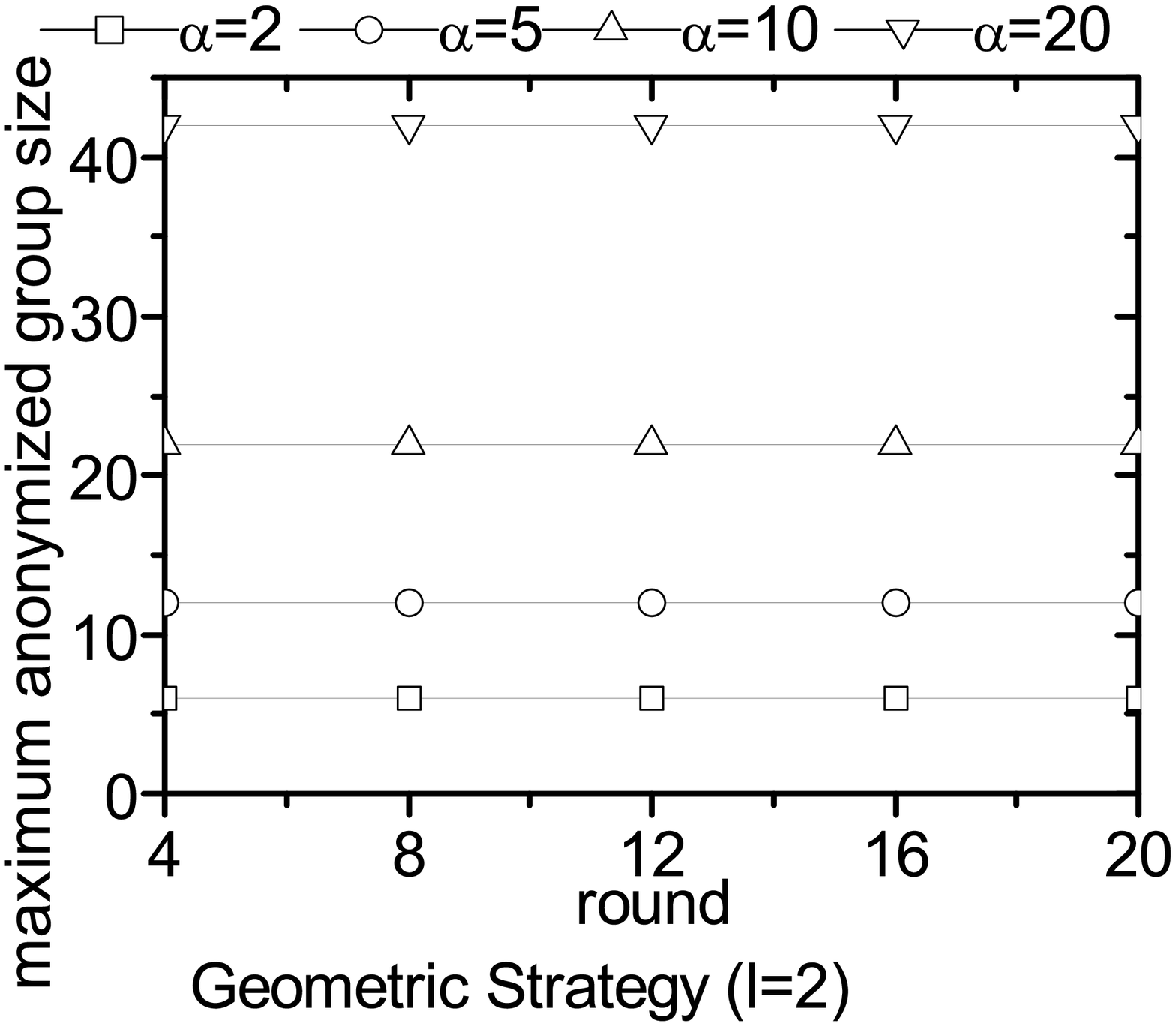}
    \end{minipage}
&
    \begin{minipage}[htbp]{4.2cm}
        \includegraphics[width=4.2cm,height=3.1cm]{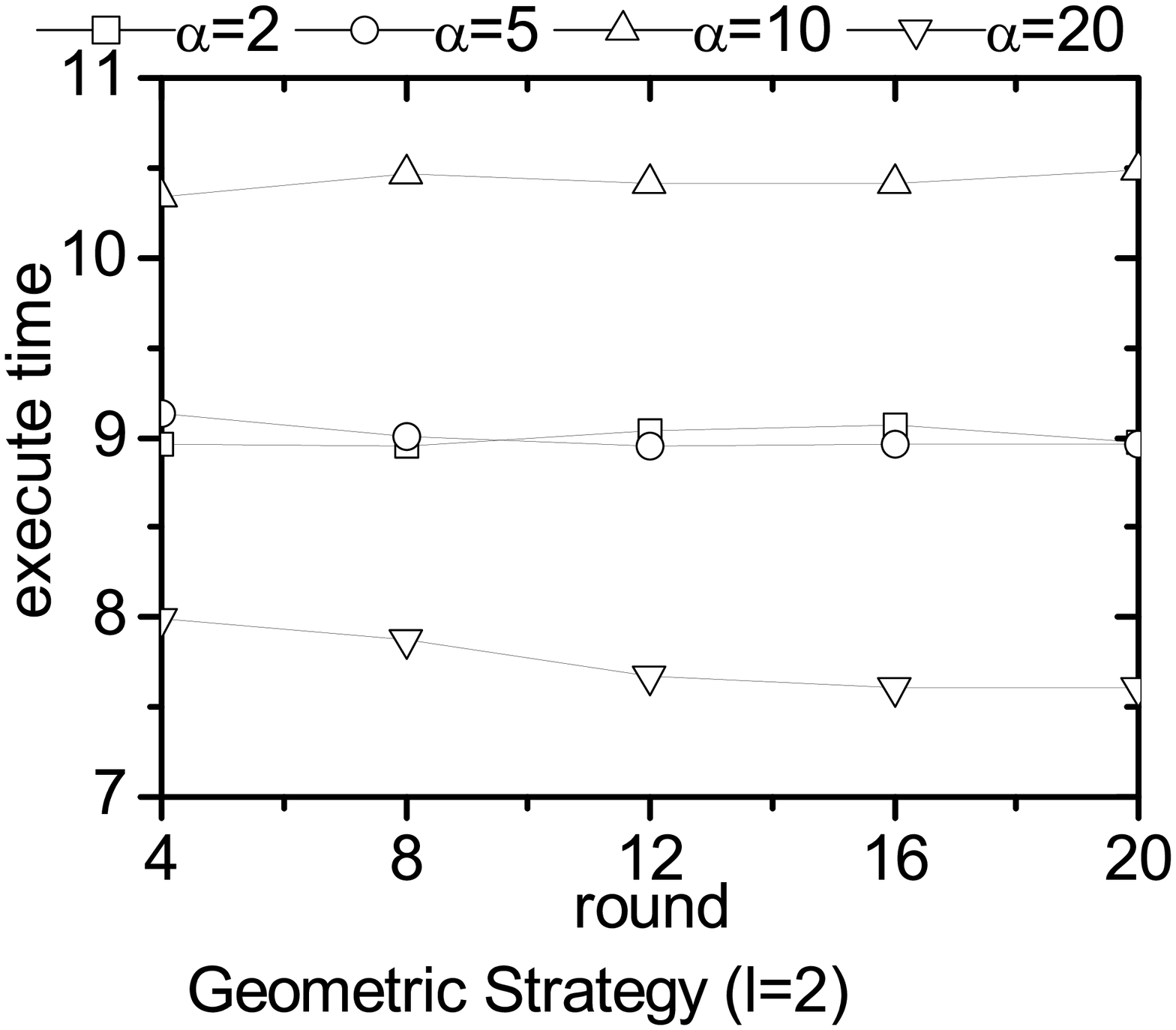}
    \end{minipage}
    \\
(a) &
 (b)
& (c) & (d)
\end{tabular}\vspace{-2mm}
\caption{Effect of $\alpha$ (Geometric Strategy) where $\ell =
2$}\vspace{-2mm}\label{fig:graphGeometricForAlpha}
\end{figure*}

 All of our experiments have been performed on a Linux
workstation with a 3.2Ghz CPU and 2 Giga-byte memory. Similar to
\cite{BFW+08}, we deploy one public available real hospital database
CADRMP\footnote{\scriptsize
http://www.hc-sc.gc.ca/dhp-mps/medeff/databasdon/index\_e.html}. In
the database, there are 8 tables: \emph{Reports}, \emph{Reactions},
\emph{Drugs}, \emph{ReportDrug}, \emph{Ingredients}, \emph{Outcome},
and \emph{Druginvolve}. \emph{Reports} consists of some patients'
basic personal information. Therefore, we take it as the voter
registration list. \emph{Reactions} has a foreign key \emph{PID}
referring to the attribute \emph{ID} in \emph{Reports} and another
attribute to indicate the person's disease. After removing tuples
with missing values, \emph{Reactions} has 105,420 tuples while
\emph{Reports} contains 40,478 different individuals.
We take 10\% least frequent sensitive values
as transient sensitive values. There are totally 232 transient
sensitive values.

Dynamic microdata table series $TS_{exp}$=\{$T_1$, $T_2$, ...,
$T_{20}$\} is created from \emph{Reactions}.
We divide \emph{Reactions} into 20 partitions of the same size, namely $P_1, P_2, ...,
P_{20}$. $T_1$ is set to $P_1$. For each $i \in [2, 20]$,
we generate $T_i$ as follows. $T_i$ is set to $P_i$ initially.
Then, we randomly select 20\% of tuples in $T_{i-1}$
and insert them into $T_i$. Then, in the resulting $T_i$,
we randomly select 20\% of tuples and change their
values in the sensitive attribute according
to the sensitive value distribution of all tuples in
\emph{Reaction} as follows.
For each selected tuple $t$ in the above step,
we randomly pick a tuple $t'$ in the
original data \emph{Reactions} and
set the sensitive value of $t$ in $T_i$
to be the sensitive value of $t'$ obtained in
\emph{Reaction}.

For our experiments, we have chosen a bottom-up anonymization
algorithm \cite{WangKe-bottom-up} with a variation of involving the
individuals that are present in the registration voter list but
absent in the data release. Such individuals can be virtually
included in an anonymized group and help to dilute the linkage
probability of individuals to sensitive values in the group \cite{NAC07,BFW+08}. This
variation helps to improve the utility since a group can now consist
of fewer records that are actually present in the data. 
A bottom-up
approach is chosen because we find that typically the anonymized
groups can be easily formed based on the smallest group sizes that
satisfy the required group size ratios. In the constant-ratio
strategy, the default value of $k'$ is equal to 20. In the geometric
strategy, the default value of $\alpha$ is set to 2.

%
%

We have tested our proposed method in terms of effectiveness and
efficiency. 
For the evaluation of our method, we examine four different aspects:
the average size of the anonymized groups in the published tables,
the greatest size of the anonymized groups in the published tables,
the utility of the published tables and the computation overheads.

For measuring the utility of the published data we compare query
processing results on each anonymized table $T_j^*$ and its
corresponding microdata table $T_j$ at each publishing round. We
follow the literature conventions~\cite{XT06b, XT07, WFW+07, BFW+08}
to measure the error by the relative error ratio in answering an
aggregate query. All the published tables are evaluated one by one.
For each evaluation, we perform 5,000 randomly generated range
queries which follows the methodology in \cite{XT07} on the
microdata snapshot and its anonymized version, and then report the
average relative error ratio.

We study the effect of variations in (1) the number of rounds, (2)
the privacy requirement $\ell$, (3) the parameter $k'$ used in the
constant-ratio strategy and (4) the parameter $\alpha$ used in the
geometric strategy.

\smallskip
\noindent\textbf{Effect of $\ell$:}
Figure~\ref{fig:graphConstantForL}(a) shows that the average
relative error of the constant-ratio strategy remains nearly
unchanged when the number of rounds (or table releases) increases.
As expected the error is larger with larger values of $\ell$. In
Figures~\ref{fig:graphConstantForL}(b) and (c), both the average
anonymized group size and the maximum anonymized group size of the
constant-ratio strategy keep nearly unchanged when we vary the
number of rounds. Again as expected, the sizes increase with $\ell$.
Figure~\ref{fig:graphConstantForL}(d) shows that the execution time
of the constant-ratio strategy keeps unchanged when there are more
rounds. In the figure, when $\ell$ is larger, the execution time is
larger. This is because we have to generate a larger anonymized
group.

Figure~\ref{fig:graphGeometricForL} shows similar results for the
geometric strategy with variation on the number of rounds. From
Figures~\ref{fig:graphGeometricForL}(a), (b) and (c) show that
the error, the average anonymized group and the maximum anonymized
group remains nearly unchanged when the number of rounds increases.
In
Figure~\ref{fig:graphGeometricForL}(d), we cannot see a consistent
trend when we vary $\ell$. The execution time when $\ell = 3$ is the
smallest. However, the execution time when $\ell = 7$ is smaller
than that when $\ell = 5$. The execution time of the algorithm
depends on two factors, namely the number of anonymized groups in
the released tables and the sizes of the anonymized groups.
Generating anonymized groups with larger sizes will increase the
execution time. On the other hand, generating fewer anonymized
groups will reduce the execution time. When $k' = 7$, since the
factor of the total number of anonymized groups (i.e., fewer
anonymized groups) outweighs the factor of the size of the
anonymized group (i.e., larger anonymized group size), the execution
time is smaller (compared with the case when $\ell = 5$).

\smallskip
\noindent\textbf{Effect of $k'$:}
We study the input parameter of $k'$ used in the constant-ratio strategy.
In Figure~\ref{fig:graphConstantForNewK}, the average relative
error, the average anonymized group size, the maximum group size
and the execution time remains nearly unchanged when the number
of rounds increases.
The average relative error, the average anonymized group size
and the maximum group size increases when $k'$ increases
as shown in Figures~\ref{fig:graphConstantForNewK}(a), (b) and (c).
In Figure~\ref{fig:graphConstantForNewK}(d), we cannot
observe a consistent trend of the execution time when $k'$ increases.
The reason is similar.

\smallskip
\noindent\textbf{Effect of $\alpha$:} We also study the input
parameter $\alpha$ for the geometric strategy. Similarly,
Figure~\ref{fig:graphGeometricForAlpha} shows that the number of
rounds does not have a significant impact on the average relative
error, the average anonymized group size, the maximum anonymized
size and the execution time.
Figures~\ref{fig:graphGeometricForAlpha}(a), (b) and (c) show that,
when $\alpha$ increases, the average relative error, the average
anonymized group size and the maximum anonymized size increases.
There is no consistent trend for the execution time when we vary
$\alpha$ as shown in Figures~\ref{fig:graphGeometricForAlpha}(d).
The reasons are similar to that in the study with the effect of
$\ell$.

Overall, our proposed methods are very efficient and introduce very
small querying error. It shows that our method can provide the
global guarantee on individual privacy as well as maintain high
utility in the published data.

%
%
%

%
%
%

\section{Conclusion}
\label{sec:concl}

In this paper, we propose a new criterion of \emph{global guarantee}
for privacy preserving data publishing. This guarantee corresponds
to a basic requirement of individual privacy where the probability
of linking an individual to a sensitive value in one or more data
releases is bounded. We show that global guarantee is a stronger
privacy requirement than localized guarantee which has been adopted
in previous works. We derive some theoretical results on this
problem and discover that the anonymized group size is an important
factor in privacy protection. According to the anonymized group
sizes, we propose two strategies for anonymization. Our empirical
study shows that these techniques are highly feasible and generate
data publication of high utility.

There are some promising future directions. In this paper, we study
the global guarantee for transient sensitive values, meaning that
the values can change freely. As a future plan, we will study the
global guarantee when both transient sensitive values and permanent
sensitive values are present. Permanent sensitive values are studied
in \cite{BFW+08} and refer to values that will be permanently linked
to an individual once it is linked to that individual. Intuitively,
we can combine the technique here and that in \cite{BFW+08} by
forming the HD-compositions for holders and decoys, as well as
forming anonymized groups based on the proper group size determined
by our strategies here for taking care of the transient values.
However, the details are left for future studies. Another direction
is to extend the problem with the consideration of other background
knowledge.

{ \small
\bibliography{transPrivacy}

\begin{thebibliography}{10}

\bibitem{EDBT04}
C.~C. Aggarwal and P.~S. Yu.
\newblock A condensation approach to privacy preserving data mining.
\newblock In {\em EDBT}, 2004.

\bibitem{AggarwalICDT05}
G.~Aggarwal, T.~Feder, K.~Kenthapadi, R.~Motwani, R.~Panigrahy, D.~Thomas, and
  A.~Zhu.
\newblock Anonymizing tables.
\newblock In {\em ICDT}, 2005.

\bibitem{Bayardo-optimal}
R.~Bayardo and R.~Agrawal.
\newblock Data privacy through optimal k-anonymization.
\newblock In {\em ICDE}, 2005.

\bibitem{BOY+05}
E.~Bertino, B.C. Ooi, Y.~Yang, and R.~Deng.
\newblock Privacy and ownership preserving of outsourced medical data.
\newblock In {\em ICDE}, 2005.

\bibitem{BFW+08}
Y.~Bu, A.~W.-C. Fu, R.~C.-W. Wong, L.~Chen, and J.~Li.
\newblock Privacy preserving serial data publishing by role composition.
\newblock In {\em VLDB}, 2008.

\bibitem{ByunSBL06}
J.~Byun, Y.~Sohn, E.~Bertino, and N.~Li.
\newblock Secure anonymization for incremental datasets.
\newblock In {\em Secure Data Management}, pages 48--63, 2006.

\bibitem{DXTZZ07}
Y.~Du, T.~Xia, Y.~Tao, D.~Zhang, and F.~Zhu.
\newblock On multidimensional k-anonymity with local recoding generalization.
\newblock In {\em ICDE}, 2007.

\bibitem{FWFP+08}
B.~C.~M. Fung, K.~Wang, A.~Fu, and J.~Pei.
\newblock Anonymity for continuous data publishing.
\newblock In {\em EDBT}, 2008.

\bibitem{GTK08}
G.~Ghinita, Y.~Tao, and P.~Kalnis.
\newblock On the anonymization of sparse high-dimensional data.
\newblock In {\em ICDE}, 2008.

\bibitem{Iyengar-kdd02}
V.~S. Iyengar.
\newblock Transforming data to satisfy privacy constraints.
\newblock In {\em KDD}, 2002.

\bibitem{KG06}
D.~Kifer and J.~Gehrke.
\newblock Injecting utility into anonymized datasets.
\newblock In {\em SIGMOD}, 2006.

\bibitem{multidimensional-Kanonymity}
K.~LeFevre, D.~DeWitt, and R.~Ramakrishnan.
\newblock Mondrian multidimensional k-anonymity.
\newblock In {\em ICDE}, 2006.

\bibitem{Incognito}
K.~LeFevre, D.~J. DeWitt, and R.~Ramakrishnan.
\newblock Incognito: Efficient full-domain k-anonymity.
\newblock In {\em SIGMOD}, 2005.

\bibitem{LL07}
N.~Li and T.~Li.
\newblock $t$-closeness: Privacy beyond $k$-anonymity and $l$-diversity.
\newblock In {\em ICDE}, 2007.

\bibitem{LL08}
T.~Li and N.~Li.
\newblock Injector: Mining background knowledge for data anonymization.
\newblock In {\em ICDE}, 2008.

\bibitem{l-diversity}
A.~Machanavajjhala, J.~Gehrke, and D.~Kifer.
\newblock $l$-diversity: privacy beyond $k$-anonymity.
\newblock In {\em ICDE}, 2006.

\bibitem{Meyerson-pods04}
A.~Meyerson and R.~Williams.
\newblock On the complexity of optimal k-anonymity.
\newblock In {\em PODS}, 2004.

\bibitem{NAC07}
M.~Nergiz, M.~Atzori, and C.W. Clifton.
\newblock Hiding the presence of individuals from shared databases.
\newblock In {\em SIGMOD}, 2007.

\bibitem{PXW+07}
J.~Pei, J.~Xu, Z.~Wang, W.~Wang, and K.~Wang.
\newblock Maintaining k-anonymity against incremental updates.
\newblock In {\em SSDBM}, 2007.

\bibitem{RSH07}
V.~Rastogi, D.~Suciu, and S.~Hong.
\newblock The boundary between privacy and utility in data publishing.
\newblock In {\em VLDB}, 2007.

\bibitem{sweeney-kanonymity-model}
L.~Sweeney.
\newblock k-anonymity: a model for protecting privacy.
\newblock {\em International journal on uncertainty, Fuzziness and knowldege
  based systems}, 10(5), 2002.

\bibitem{TXLZ08}
Y.~Tao, X.~Xiao, J.~Li, and D.~Zhang.
\newblock On anti-corruption privacy preserving publication.
\newblock In {\em ICDE}, 2008.

\bibitem{WF+06}
K.~Wang and B.~C.~M. Fung.
\newblock Anonymizing sequential releases.
\newblock In {\em EDBT}, 2008.

\bibitem{WangKe-template}
K.~Wang, B.~C.~M. Fung, and P.~S. Yu.
\newblock Template-based privacy preservation in classification problems.
\newblock In {\em ICDM05}, 2005.

\bibitem{WangKe-bottom-up}
K.~Wang, P.~S. Yu, and S.~Chakraborty.
\newblock Bottom-up generalization: A data mining solution to privacy
  protection.
\newblock In {\em ICDM}, 2004.

\bibitem{WFW+07}
R.C.W. Wong, A.~Fu, K.~Wang, and J.~Pei.
\newblock Minimality attack in privacy preserving data publishing.
\newblock In {\em VLDB}, 2007.

\bibitem{WLFW-kdd06}
R.C.W. Wong, J.~Li, A.~Fu, and K.~Wang.
\newblock (alpha, k)-anonymity: An enhanced k-anonymity model for
  privacy-preserving data publishing.
\newblock In {\em KDD}, 2006.

\bibitem{XT06b}
X.~Xiao and Y.~Tao.
\newblock Anatomy: Simple and effective privacy preservation.
\newblock In {\em {VLDB}}, 2006.

\bibitem{XT06a}
X.~Xiao and Y.~Tao.
\newblock Personalized privacy preservation.
\newblock In {\em SIGMOD}, 2006.

\bibitem{XT07}
X.~Xiao and Y.~Tao.
\newblock $m$-invariance: Towards privacy preserving re-publication of dynamic
  datasets.
\newblock In {\em SIGMOD}, 2007.

\bibitem{XWP-kdd06}
J.~Xu, W.~Wang, J.~Pei, X.~Wang, B.~Shi, and A.~Fu.
\newblock Utility-based anonymization using local recoding.
\newblock In {\em KDD}, 2006.

\bibitem{XWFY08}
Y.~Xu, K.~Wang, A.~W.-C. Fu, and P.S. Yu.
\newblock Anonymizing transaction databases for publication.
\newblock In {\em KDD}, 2008.

\bibitem{ZKS+07}
Q.~Zhang, N.~Koudas, D.~Srivastava, and T.~Yu.
\newblock Aggregate query answering on aononymized tables.
\newblock In {\em ICDE}, 2007.

\end{thebibliography}
}

\normalsize
\section{Appendix}

\vspace*{3mm}

Here we give the proofs of some of the lemmas and theorems listed in
the previous sections.

\normalsize

Theorem \ref{lemma:setMinimumSizePrivacyBreach-new}: \emph{Consider
that we published $k-1$ tables where an equivalence class in
$T_{k-1}^*$ containing $t$ is linked to $s_1$. Suppose we are to
publish $T_k^*$ where an equivalence class in $T_k^*$ containing $o$
is also linked to $s_1$. If $\frac{n_{k-1}}{n_{k-1,1}} =
\widetilde{\underline{n}}(k-1)$, then $p(o,s_1,k) > 1/\ell$.}

\medskip

\textbf{Proof:}
\begin{eqnarray*}
\frac{n_{k-1}}{n_{k-1,1}} & = & \widetilde{\underline{n}}(k-1)\\
    & = & \frac{\ell \prod_{j = 1}^{k-2}(n_j - n_{j, 1})}{\ell \prod_{j=1}^{k-2}(n_j - n_{j, 1}) - (\ell -1)\prod_{j=1}^{k-2}n_j}\\
    & = & \frac{\ell \prod_{j=1}^{k-1}(1 - \frac{n_{j, 1}}{n_j})}{\ell \prod_{j=1}^{k-2}(1- \frac{n_{j,1}}{n_j}) - (\ell -1)}
\end{eqnarray*}
Let $P = \prod_{j=1}^{k-1}(1 - \frac{n_{j, 1}}{n_j})$.
We have
$$
\frac{n_{k-1}}{n_{k-1,1}} = \frac{\ell P}{\ell P - \ell + 1}
$$
Consider
\begin{eqnarray*}
  &  &  p(o, s_1, k) \\
  & = & \frac{\prod_{j=1}^{k}n_j - \prod_{j=1}^{k}(n_j - n_{j, 1})}{\prod_{j=1}^{k}n_j}\\
  & = & 1 - \prod_{j=1}^{k} (1 - \frac{n_{j, 1}}{n_j}) \\
  & = & 1 - (1 - \frac{n_{k,1}}{n_k}) ( 1 - \frac{n_{k-1,1}}{n_{k-1}}) \prod_{j=1}^{k-2}(1 - \frac{n_{j,1}}{n_j})\\
  & = & 1 - (1 - \frac{n_{k,1}}{n_k}) ( 1 - \frac{\ell P}{\ell P - \ell + 1}) P\\
  & = & 1 - (1 - \frac{n_{k,1}}{n_k}) (  1 - \frac{1}{\ell})\\
  & = & \frac{1}{\ell} + \frac{n_{k,1}}{n_k}(1-\frac{1}{\ell})\\
  & > & \frac{1}{\ell}
\end{eqnarray*}
That is, $p(o,s_1,k) > 1/\ell$. \done

\bigskip

Theorem \ref{lemma:monotonicity} ({\small MONOTONICITY}).
\emph{$p(o,s_1,k)$ is strictly decreasing when $\frac{n_k}{n_{k,1}}$
increases.}

\medskip

\textbf{Proof:} 
\begin{eqnarray*}
 p(o, s_1, k) & = & \frac{\prod_{j=1}^k n_j - \prod_{j=1}^k (n_j - n_{j,1})}{\prod_{j=1}^{k} n_j}\\
              & = & 1 - \prod_{j=1}^k (1 - \frac{n_{j,1}}{n_j})\\
              & = & 1 - (1 - \frac{n_{k,1}}{n_k})\prod_{j=1}^{k-1} (1 - \frac{n_{j,1}}{n_j})
\end{eqnarray*}
If $\frac{n_k}{n_{k,1}}$ increases, the above equation decreases.
\done

\end{sloppy}

\end{document}